\documentclass[nofootinbib,aps,11pt]{revtex4}
\usepackage{graphicx}
\usepackage{cancel}
\usepackage{amssymb}
\usepackage{textcomp}
\usepackage{amsmath}
\usepackage{bm}
\usepackage{times}
\usepackage{epsfig}
\usepackage{color}
\usepackage{axodraw}
\usepackage{graphicx}

\def\lsim{\:\raisebox{-0.5ex}{$\stackrel{\textstyle<}{\sim}$}\:}

\def\half{\textstyle{1\over2}}

\begin{document}

\title{\Large Neutrino Masses in Split Supersymmetry}
\author{Marco Aurelio D\'\i az$^{1}$}
\author{Pavel Fileviez P\'erez$^{2,3}$}
\author{Clemencia Mora$^{4}$}
\affiliation{
{\small $^{1}$ Departamento de F\'\i sica, Universidad Cat\'olica de Chile,
Avenida Vicu\~na Mackenna 4860, Santiago, Chile 
\\
$^{2}$ Department of Physics, University of Wisconsin-Madison, 
1150 University Avenue, Madison, WI 53706, USA
\\
$^{3}$ CFTP, Instituto Superior Tecnico, Ave. Rovisco Pais 1,
1049-001, Lisbon, Portugal
\\
$^{4}$ D\'epartement de Physique Nucl\'eaire et Corpusculaire ,
Universit\'e de Gen\`eve,  24 Quai Ernest-Ansermet, 1211 Gen\`eve 4,
Suisse
}}
\begin{abstract}
We investigate the possibility to generate neutrino masses in the 
context of Split supersymmetric scenarios where all sfermions are 
very heavy. All relevant contributions coming from the R-parity violating 
terms to the neutrino mass matrix up to one-loop level are computed, 
showing the importance of the Higgs one-loop corrections. We conclude 
that it is not possible to generate all neutrino masses 
and mixings in Split SUSY with bilinear R-Parity violating interactions. 
In the case of Partial Split SUSY the one-loop Higgs contributions 
are enough to generate the neutrino masses and mixings in agreement 
with the experiment. In the context of minimal SUSY SU(5) we find 
new contributions which help us to generate neutrino masses 
in the case of Split SUSY.  
\end{abstract}
\date{\today}
\maketitle
\section{Introduction}
Supersymmetric extensions of the Standard Model (SM) have been 
considered as one of the most appealing candidates for physics beyond 
the Standard Model (SM). Recently, different supersymmetric scenarios 
have been studied extensively. We mention low-energy SUSY~\cite{Haber:1984rc}, 
where the supersymmetric scale is around TeV, and Split SUSY where 
all the scalars, except for one Higgs doublet, are very heavy~\cite{SPLIT}.
In both supersymmetric scenarios mentioned above it is possible to achieve 
unification of the gauge interactions at the high scale and the lightest 
supersymmetric particle (LSP) could be a natural candidate to describe 
the Cold Dark Matter of the Universe once the so-called R-parity 
is imposed as an exact symmetry of the theory. In SPLIT SUSY scenarios, 
ignoring the hierarchy problem, most of the unpleasant aspects of 
low-energy SUSY, such as excessive flavour and CP violation, 
and very fast dimension 5 proton decay, are eliminated.   

It is very-well known that in general interactions which break the 
lepton or baryon number (or R-parity) are present in any SUSY extension 
of the SM. Therefore, we have the possibility to generate the 
neutrino masses and mixing~\cite{NeutrinoReview}, and we have to understand 
the predictions for proton stability~\cite{ProtonReview}. 
For several phenomenological aspects of R-parity violating interactions 
see Ref.~\cite{Rparity}. The possibility to describe the neutrino 
properties with R-parity violating interactions in the context of the 
Minimal Supersymmetric Standard Model has been studied in detail 
by several groups in the context of low energy supersymmetry~
(See for example Refs.~\cite{Hirsch:2000ef} and 
\cite{Rparity-neutrinos}). In the context of SPLIT SUSY the possibility to 
describe the masses and mixing of neutrinos has been studied in 
Ref.~\cite{Chun:2004mu}, where the authors concluded that it is not 
possible to use the R-parity bilinear terms alone to describe the neutrino 
properties.
 
In this work we re-examine the possibility to describe the properties of 
neutrinos using the R-parity violating interactions in the context of split 
supersymmetric scenarios. We agree with the results presented in 
Ref.~\cite{Chun:2004mu} that in Split Supersymmetry, where only one 
Higgs doublet remains at the weak scale, it is not possible to generate 
the neutrino masses in agreement with the experiments and explain the 
reasons in detail. We study an alternative Split SUSY scenario where only 
the sfermions are very heavy while all Higgses can be light. We refer to 
this scenario as ``Partial SPLIT SUSY''. Notice that in this scenario we 
can keep the nice features of SPLIT SUSY such as the suppression of proton 
decay due to R-parity violation and unification of gauge couplings at the 
high-scale. In this SUSY scenario we show that it is possible to generate 
the neutrino masses using all relevant interactions once the heavy sfermions 
are integrated out. Computing all contributions up to one-loop level, we find 
an example solution where it is shown that all constraints coming from neutrino 
experiments on the R-parity violating interactions are satisfied. In this 
scenario even if R-Parity is broken one could have the gravitino 
as a possible cold dark matter candidate.

We conclude that in Partial SPLIT SUSY it is possible to generate all 
neutrino masses and mixing in agreement with the experiments using the 
bilinear terms alone and the trilinear R-Parity violating (TRpV) couplings 
are essentially irrelevant. The key element is that the symmetry of the 
neutrino mass matrix at tree level is broken by the Higgs bosons loops 
together with neutralinos and charginos. The terms that break the symmetry 
of neutrino mass matrix vanish in the decoupling limit, making not possible 
the description of the neutrino masses in the ``Standard'' Split SUSY 
scenario. We study the same issue in the context of the minimal 
supersymmetric SU(5) where one finds new contributions which help 
us to generate neutrino masses in agreement with the experiments 
in the case of SPLIT SUSY. 
\section{R-parity violation and neutrino masses in SPLIT SUSY}
As we know in any supersymmetric extension of the Standard Model 
there are interactions terms which break the so-called R-parity. 
The R-parity is defined as $R=(-1)^{3(B-L)+ 2S}$, where 
$L$, $B$, and $S$ are the lepton and baryon number, and the spin, respectively. 
Usually this symmetry is considered as an exact symmetry of the minimal 
supersymmetric extension of the Standard Model (MSSM) in order to avoid 
the dimension four contributions to proton decay and at the same time 
there is a possibility to have the lightest supersymmetric particle 
as a good candidate for the Cold Dark Matter of the Universe.

In the context of the MSSM the so-called R-parity violating 
terms are given by
\begin{equation}
{\cal W}_{NR}= \alpha_{ijk} \hat{Q}_i \hat{L}_j \hat{D}^C_k \ + \ 
\beta_{ijk} \hat{U}_i^C \hat{D}_j^C \hat{D}^C_k \ + \
\gamma_{ijk} \hat{L}_i \hat{L}_j \hat{E}^C_k \ + 
\ \epsilon_i \hat{L}_i \hat{H}_u,
\end{equation}
where $\beta_{ijk}=-\beta_{ikj}$ and $\gamma_{ijk}=-\gamma_{jik}$. As it is 
well-known due to the presence of the first and second terms in the above 
equation one has the so-called the dimension four contributions to the 
decay of the proton. In this case in order to satisfy the experimental 
bounds on the proton decay lifetime one has to assume that the 
multiplication of the couplings $\alpha_{ijk}$ and $\beta_{ijk}$ is 
of the order $10^{-21}$ when the susy scale is at electroweak scale. 
In order to avoid these very small couplings in the theory one 
imposes by hand the R-parity symmetry. There is a second way to 
avoid these small couplings if the susy breaking scale is large, 
this is the case of Split-SUSY. Since in this case there is no 
need to impose any symmetry by hand we stick to this possibility 
and study the generation of neutrino masses in this context.    

Let us discuss how to generate neutrino masses through 
this mechanism in three different scenarios:

\begin{itemize}

\item {\textit{\bf MSSM with SPLIT SUSY}}

In this supersymmetric scenario called Split SUSY all scalars are very 
heavy, except for one Higgs doublet. Integrating out the heavy scalars 
all possible R-parity conserving interactions in split 
supersymmetric scenarios are given by~\cite{SPLIT}:
\begin{eqnarray}
{\cal L}^{split}_{susy}&=& {\cal L}^{split}_{kinetic} \ + \ m^2H^\dagger 
H - \frac{\lambda}{2}(H^\dagger H)^2 
-\Big[ Y_u \overline q_L u_R i \sigma_2 H^* \ + \ Y_d \overline q_L d_R H \ 
+ \ Y_e \overline l_L e_R H \ + \
\nonumber\\ 
&&
+ \frac{M_3}{2} \widetilde G\widetilde G \ + \ \frac{M_2}{2} \widetilde W 
\widetilde 
W \ + \ \frac{M_1}{2} \widetilde B \widetilde B \ + \ 
\mu \widetilde H_u^T i \sigma_2 \widetilde H_d \ + \
\label{LagSplit}\\ &&
+\textstyle{\frac{1}{\sqrt{2}}} H^\dagger
(\tilde g_u \sigma\widetilde W \ + \ \tilde g'_u\widetilde B)\widetilde H_u
 \ + \ \textstyle{\frac{1}{\sqrt{2}}} H^T i \sigma_2
(-\tilde g_d \sigma \widetilde W+ \tilde g'_d \widetilde B)\widetilde H_d 
+\mathrm{h.c.}\Big],
\nonumber
\end{eqnarray}
where
\begin{equation}
H=\left(\begin{matrix} H^+ 
\cr \frac{1}{\sqrt{2}} (v+\phi^0+i\varphi^0) \end{matrix}\right),
\end{equation}
is the SM Higgs. In the above equations we have the SM fields 
$q_L$, $u_R$, $d_R$, $l_L$, $e_R$ and the superpartners 
of the Higgs and gauge bosons present in the MSSM. Following our notation 
$\widetilde G$, $\widetilde W$, and $\widetilde B$ are the gauginos 
associated to the $SU(3)$, $SU(2)$, and $U(1)$ gauge groups, respectively. 
While $\widetilde H_u$ and $\widetilde H_d$ correspond to the up and down 
higgsinos. The parameters in Eq.~(\ref{LagSplit}) are the following: 
$m$ is the Higgs mass parameter, $\lambda$ is the Higgs quartic self 
coupling; $Y_u$, $Y_d$, and $Y_e$ are the Yukawa couplings; $M_3$, $M_2$, 
and $M_1$ are the gaugino masses, $\mu$ the higgsino mass, 
and $\tilde g_u$, $\tilde g'_u$, $\tilde g_d$, and $\tilde g'_d$ 
are trilinear couplings between the Higgs boson, gauginos, and higgsinos.

The Higgs-gaugino-higgsino couplings in Eq.~(\ref{LagSplit}) satisfy 
matching conditions at the scale $\tilde m$. Above this scale, the 
theory is supersymmetric and the squarks, sleptons, and heavy Higgs 
doublet have a mass assumed to be nearly degenerate equal to $\tilde m$. 
The supersymmetric lagrangian includes the terms,
\begin{equation}
{\cal L}_{susy}\owns-\mu\widetilde H^T_ui\sigma_2\widetilde H_d
-\frac{H_u^\dagger}{\sqrt{2}}\left(
g\sigma\widetilde W+g'\widetilde B\right)\widetilde H_u
-\frac{H_d^\dagger}{\sqrt{2}}\left(
g\sigma\widetilde W - g'\widetilde B\right)\widetilde H_d,
\end{equation}
which implies the following boundary conditions at $\tilde m$:
\begin{eqnarray}
\tilde g_u(\tilde m)=g(\tilde m) \ \sin\beta(\tilde m)
&\,,\qquad&
\tilde g_d(\tilde m)=g(\tilde m) \ \cos\beta(\tilde m),
\nonumber\\
\tilde g'_u(\tilde m)=g'(\tilde m) \ \sin\beta(\tilde m)
&\,, \text{and} \qquad &
\tilde g'_d(\tilde m)=g'(\tilde m) \ \cos\beta(\tilde m),
\end{eqnarray}
where $g(\tilde m)$ and $g'(\tilde m)$ are the gauge coupling constants 
evaluated at the scale $\tilde m$. At the same time the angle $\beta$ 
is the mixing angle between the two Higgs doublets $H_d$ and $H_u$ 
of the supersymmetric model. In order to set our notation 
the two doublets are given by,
\begin{equation}
H_d=\left(\begin{matrix} \frac{1}{\sqrt{2}} (v_d+\phi^0_d+i\varphi^0_d)
\cr H_d^- \end{matrix}\right),
\,\qquad
H_u=\left(\begin{matrix} H_u^+ 
\cr \frac{1}{\sqrt{2}} (v_u+\phi^0_u+i\varphi^0_u) \end{matrix}\right),
\end{equation}
and $\tan\beta=v_u/v_d$. In terms of these two Higgs doublets of the MSSM, 
the light fine-tuned Higgs doublet $H$ in the low energy effective model is 
$H=-i\sigma_2 H^*_d cos \beta(\tilde m)+H_u sin \beta(\tilde m)$. 

As we mentioned before in SPLIT SUSY scenarios at low energy 
we have the SM fields, the charginos and neutralinos. Using the 
above notation the chargino mass matrix is given by:
\begin{equation}
{\bf M}_{\chi^+}^{SS}=\left[\begin{array}{cc}
M_2 & \frac{1}{\sqrt{2}} \tilde g_u v \\
\frac{1}{\sqrt{2}} \tilde g_d v & \mu
\end{array}\right],
\label{chamassmat}
\end{equation}
while the neutralino mass matrix reads as:
\begin{equation}
{\bf M}_{\chi^0}^{SS}=\left[\begin{array}{cccc}
M_1 & 0 & -\frac{1}{2}\tilde g'_d v & \frac{1}{2}\tilde g'_u v \\
0 & M_2 & \frac{1}{2}\tilde g_d v & -\frac{1}{2}\tilde g_u v \\
-\frac{1}{2}\tilde g'_d v & \frac{1}{2}\tilde g_d v & 0 & -\mu \\
\frac{1}{2}\tilde g'_u v & -\frac{1}{2}\tilde g_u v & -\mu & 0
\end{array}\right].
\label{X0massmat}
\end{equation}
Now, since we are interested in the possibility to describe the 
neutrino masses 
in Split-SUSY, we write all relevant R-Parity violating interactions:
\begin{equation}
{\cal L}^{split}_{RpV}=
\epsilon_i \widetilde H_u^T i \sigma_2 L_i \ -\ 
\textstyle{\frac{1}{\sqrt{2}}} a_i H^T i \sigma_2
(-\tilde g_d \sigma\widetilde W+\tilde g'_d\widetilde B)L_i \ + \ h.c., 
\label{LSplitRpV}
\end{equation}
where $\epsilon_i$ are the parameters that mix higgsinos with
leptons, and $a_i$ are dimensionless parameters that mix gauginos with
leptons. Notice that the first term is the usual bilinear term, 
while the last two terms are obtained once we integrate 
out the sleptons using the bilinear soft terms ($\tilde{L}_i H_u$) which 
break explicitly R-parity. As it is well-known we can also write 
the usual R-parity violating trilinear terms 
($\hat{Q} \hat{D}^C \hat{L}$, $\hat{L} \hat{L} \hat{E}^C$). 
However, since the sfermions are very heavy in SPLIT SUSY 
and the contributions to the neutrino mass matrix 
coming from those terms are at one-loop level, 
those interactions cannot play any important role. 
Using Eq.~(\ref{LSplitRpV}), after the Higgs acquires a vev, 
we find the relevant terms for neutrino masses:
\begin{equation}
{\cal L}^{split}_{RpV} = -\left[
\epsilon_i \widetilde H_u^0 + \frac{1}{2} a_i v \left( 
\tilde g c_\beta\widetilde W_3 - \tilde g' c'_\beta\widetilde B \right)
\right] \nu_i \ + \ h.c. \ + \ \ldots,
\end{equation}
where $v$ is the vacuum expectation value of the SM-like Higgs field $H$.
Knowing all R-parity violating interactions we can write 
the neutralino/neutrino mass matrix as:
\begin{equation}
{\cal M}_N^{SS}=\left[\begin{array}{cc} {\mathrm M}_{\chi^0}^{SS} & (m^{SS})^T \\ 
m^{SS} & 0 \end{array}\right],
\label{X07x7}
\end{equation}
where ${\mathrm M}_{\chi^0}^{SS}$ is given by Eq.~(\ref{chamassmat}) 
and $m^{SS}$ reads as: 
\begin{equation}
m^{SS}=\left[\begin{array}{cccc}
-\frac{1}{2} \tilde g'_d a_1v & \frac{1}{2} \tilde g_d a_1v 
& 0 &\epsilon_1 \cr
-\frac{1}{2} \tilde g'_d a_2v & \frac{1}{2} \tilde g_d a_2v&0 
& \epsilon_2 \cr
-\frac{1}{2} \tilde g'_d a_3v & \frac{1}{2} \tilde g_d a_3v&0 
& \epsilon_3
\end{array}\right].
\end{equation}
We define the parameters $\lambda_i\equiv a_i\mu+\epsilon_i$, which
are related to the traditional BRpV parameters $\Lambda_i$
\cite{Nowakowski:1995dx} by $\Lambda_i=\lambda_i v_d$.
Integrating out the neutralinos, we find that the neutrino mass matrix 
is given by:
\begin{equation}
{\bf M}_\nu^{eff}=-m^{SS}\,({\mathrm{M}}_{\chi^0}^{SS})^{-1}\,(m^{SS})^T=
\frac{v^2}{4\det{M_{\chi^0}^{SS}}}
\left(M_1 \tilde g^2_d + M_2 \tilde g'^2_d \right)
\left[\begin{array}{cccc}
\lambda_1^2        & \lambda_1\lambda_2 & \lambda_1\lambda_3 \cr
\lambda_2\lambda_1 & \lambda_2^2        & \lambda_2\lambda_3 \cr
\lambda_3\lambda_1 & \lambda_3\lambda_2 & \lambda_3^2
\end{array}\right],
\label{treenumass}
\end{equation}
where the determinant of the neutralino mass matrix is:
\begin{equation}
\det{M_{\chi^0}^{SS}}=-\mu^2 M_1 M_2 + \frac{1}{2} v^2\mu \left( 
M_1 \tilde g_u \tilde g_d + M_2 \tilde g'_u \tilde g'_d \right)
+\textstyle{\frac{1}{16}} v^4 
\left(\tilde g'_u \tilde g_d - \tilde g_u \tilde g'_d \right)^2.
\label{detNeut}
\end{equation}
Notice that the effective neutrino mass matrix ${\bf M}_\nu^{eff}$ has 
only one eigenvalue different from zero. As in the case of R-parity 
violation in the MSSM with bilinear terms, at tree level only 
one neutrino is massive. Therefore, we have to investigate all possible 
one loop contributions to the neutrino mass matrix which help us to 
generate the atmospheric and solar neutrino masses. It has been argued in the 
literature~\cite{Chun:2004mu} that using the bilinear terms it is not 
possible to explain the neutrino masses and mixing. We study this issue in 
detail and as we will show in the next section that once we include the 
one-loop contributions to the neutrino mass matrix it is not possible to 
generate all neutrino masses in agreement with the experiment.  
\item {\textit{\bf MSSM with Partial SPLIT SUSY}}

Let us study the same issue, how to generate the 
neutrino masses through the R-parity violating interactions, in 
Partial SPLIT SUSY where only the sfermions are very heavy 
while the Higgs can be light. Notice that in this case proton decay 
can be suppressed and the unification of the gauge interactions 
at the high scale is possible as well. We will show that in this 
scenario the contributions from the light Higgs bosons is enough 
to generate the neutrino masses at one-loop, and study the decoupling limit in 
order to have a better understanding of the results presented in 
the previous section.

We integrate out the heavy squarks and sleptons and find that the R-parity 
conserving (RpC) interactions below the scale $\widetilde m$ are given by
\begin{eqnarray}
{\cal L}_{PSS}^{RpC} &\owns& - \Big[ m_1^2H_d^\dagger H_d 
+ m_2^2H_u^\dagger H_u - m_{12}^2(H_d^T\epsilon H_u+h.c.)
\nonumber\\ &&
+ \textstyle{1\over2}\lambda_1(H_d^\dagger H_d)^2 
+ \textstyle{1\over2}\lambda_2(H_u^\dagger H_u)^2 
+ \lambda_3(H_d^\dagger H_d)(H_u^\dagger H_u)
+ \lambda_4|H_d^T\epsilon H_u|^2
\Big]
\nonumber\\ &&
+ h_u \overline u_R H_u^T \epsilon q_L 
- h_d \overline d_R H_d^T \epsilon q_L
- h_e \overline e_R H_d^T \epsilon l_L \ - \
\label{LagSS2HDM}\\ &&
-\textstyle{\frac{1}{\sqrt{2}}} H_u^\dagger
(\tilde g_u \sigma\widetilde W + \tilde g'_u\widetilde B)\widetilde H_u
-\textstyle{\frac{1}{\sqrt{2}}} H_d^\dagger
(\tilde g_d \sigma \widetilde W - \tilde g'_d \widetilde B)\widetilde H_d 
+\mathrm{h.c.}
\nonumber
\end{eqnarray}
In the above equations, the two Higgs doublets that survive at 
the weak scale are $H_d$ and $H_u$. The parameters 
in Eq.~(\ref{LagSS2HDM}) not defined before are the following: $m_1^2$, 
$m_2^2$, and $m_{12}^2$ are the Higgs mass parameters, 
$\lambda_i$, $i=1,2,3,4$ are the Higgs quartic self couplings; and $h_u$, 
$h_d$, and $h_e$ are the Yukawa couplings.
The Higgs-gaugino-higgsino, gauge, and Yukawa couplings in 
Eq.~(\ref{LagSS2HDM}) satisfy matching conditions at the scale $\tilde m$. 
Above this scale, the theory is supersymmetric and the squarks and sleptons
have a mass assumed to be nearly degenerate to $\tilde m$. The 
supersymmetric lagrangian above $\tilde m$ includes the terms,
\begin{eqnarray}
{\cal L}_{susy}^{RpC} &\owns& - \Big[ m_1^2H_d^\dagger H_d 
+ m_2^2H_u^\dagger H_u - m_{12}^2(H_d^T\epsilon H_u+h.c.)
+ \textstyle{1\over8}(g^2+g'^2)(H_d^\dagger H_d)^2
\nonumber\\ &&
+ \textstyle{1\over8}(g^2+g'^2)(H_u^\dagger H_u)^2
+ \textstyle{1\over4}(g^2-g'^2)(H_d^\dagger H_d)(H_u^\dagger H_u)
- \textstyle{1\over2}g^2|H_d^T\epsilon H_u|^2
\Big]
\nonumber\\ &&
+ \lambda_u \overline u_R H_u^T \epsilon q_L 
- \lambda_d \overline d_R H_d^T \epsilon q_L
- \lambda_e \overline e_R H_d^T \epsilon l_L
\label{LagSplit2}\\ &&
-\textstyle{\frac{1}{\sqrt{2}}} H_u^\dagger
(g \sigma\widetilde W + g'\widetilde B)\widetilde H_u
-\textstyle{\frac{1}{\sqrt{2}}} H_d^\dagger
(g \sigma \widetilde W - g' \widetilde B)\widetilde H_d 
+\mathrm{h.c.}
\nonumber
\end{eqnarray}
Consequently, at the scale $\tilde m$ we have the following boundary 
conditions for the Higgs couplings,
\begin{equation}
\lambda_1=\lambda_2=\textstyle{1\over4}(g^2+g'^2) \,, \qquad
\lambda_3=\textstyle{1\over4}(g^2-g'^2) \,, \qquad
\lambda_4=-\textstyle{1\over2}g^2,
\end{equation}
for the Yukawa couplings $h_u\!=\!\lambda_u$\,,\, $h_d\!=\!\lambda_d$\,,\, 
$h_e\!=\!\lambda_e$\,, and for the higgsino-gaugino Yukawa couplings, 
$\tilde g_u\!=\!\tilde g_d\!=\!g$\,,\, 
$\tilde g'_u\!=\!\tilde g'_d\!=\!g'$\,. All of them 
evaluated at the scale $\widetilde m$. Note the difference 
between these boundary conditions and the corresponding ones in the original
Split Supersymmetric model: the former do not involve the angle $\beta$.
At the weak scale, the minimization of the Higgs potential leads to a 
vacuum expectation value for both Higgs doublets which satisfy
$v_d^2+v_u^2=v^2$, such that $m_W^2=\textstyle{1\over2}g^2v^2$ and
$m_Z^2=\textstyle{1\over2}(g^2+g'^2)v^2$, as usual for a two Higgs 
doublet model (2HDM).

As we mentioned before in SPLIT SUSY scenarios, the charginos and 
neutralinos survive at low energies. Using the above notation the chargino 
mass matrix is given by:
\begin{equation}
{\bf M}_{\chi^+}^{PSS}=\left[\begin{array}{cc}
M_2 & \frac{1}{\sqrt{2}} v \tilde g_u s_\beta \\
\frac{1}{\sqrt{2}} v \tilde g_d c_\beta & \mu
\end{array}\right],
\label{chamassmat2}
\end{equation}
while the neutralino mass matrix reads as:
\begin{equation}
{\bf M}_{\chi^0}^{PSS}=\left[\begin{array}{cccc}
M_1 & 0 & -\frac{1}{2}\tilde g'_dc_\beta v & \frac{1}{2}\tilde g'_us_\beta v \\
0 & M_2 & \frac{1}{2}\tilde g_dc_\beta v & -\frac{1}{2}\tilde g_us_\beta v \\
-\frac{1}{2}\tilde g'_dc_\beta v & \frac{1}{2}\tilde g_dc_\beta v & 0 & -\mu \\
\frac{1}{2}\tilde g'_us_\beta v & -\frac{1}{2}\tilde g_us_\beta v & -\mu & 0
\end{array}\right].
\label{X0massmat2}
\end{equation}
The difference with the Split Supersymmetric case in Eqs.~(\ref{chamassmat}) 
and (\ref{X0massmat}) is in the mixings between higgsinos and gauginos.
Now, with the neutrino masses in mind, we write all relevant R-Parity 
violating interactions in Partial SPLIT SUSY:
\begin{equation}
{\cal L}_{PSS}^{RpV} =
-\epsilon_i \widetilde H_u^T \epsilon L_i  
\ -\ 
\textstyle{\frac{1}{\sqrt{2}}} b_i H_u^T\epsilon
(\tilde g_d \sigma\widetilde W-\tilde g'_d\widetilde B)L_i 
\ + \ h.c., 
\label{LSS2HDMRpV}
\end{equation}
with $b_i$ dimensionless parameters. Using Eq.~(\ref{LSS2HDMRpV}), 
after the Higgs acquires a vev, we find the relevant terms 
for neutrino masses:
\begin{equation}
{\cal L}^{RpV}_{PSS} = - \left[
\epsilon_i \widetilde H_u^0 + \frac{1}{2} b_i v_u \left( 
\tilde g_d \widetilde W_3 - \tilde g'_d \widetilde B \right)
\right] \nu_i \ + \ h.c. \ + \ \ldots,
\end{equation}
where $v_d=v c_\beta$ and $v_u=v s_\beta$ are the vev of the two Higgs 
doublets. The neutralino/neutrino mass matrix still has the form given
in Eq.~(\ref{X07x7}), but in this scenario the matrix $m$ reads as,
\begin{equation}
m^{PSS}=\left[\begin{array}{cccc}
-\frac{1}{2} \tilde g'_d b_1 v_u & 
 \frac{1}{2} \tilde g_d  b_1 v_u & 
0 &\epsilon_1 
\cr
-\frac{1}{2} \tilde g'_d b_2 v_u & 
 \frac{1}{2} \tilde g_d  b_2 v_u & 
0 & \epsilon_2 
\cr
-\frac{1}{2} \tilde g'_d b_3 v_u & 
 \frac{1}{2} \tilde g_d  b_3 v_u &
0 & \epsilon_3
\end{array}\right].
\end{equation}
The effective neutrino mass matrix obtained 
after diagonalizing by blocks is,
\begin{equation}
{\bf M}_\nu^{eff}=
-m^{PSS}\,({\mathrm{M}}_{\chi^0}^{PSS})^{-1}\,(m^{PSS})^T=
\frac{M_1 \tilde g^2_d + M_2 \tilde {g'}^2_d}{4\det{M_{\chi^0}^{PSS}}}
\left[\begin{array}{cccc}
\Lambda_1^2        & \Lambda_1\Lambda_2 & \Lambda_1\Lambda_3 \cr
\Lambda_2\Lambda_1 & \Lambda_2^2        & \Lambda_2\Lambda_3 \cr
\Lambda_3\Lambda_1 & \Lambda_3\Lambda_2 & \Lambda_3^2
\end{array}\right],
\label{treenumass2}
\end{equation}
with $\Lambda_i=\mu b_i v_u + \epsilon_i v_d$, and with the determinant 
of the neutralino submatrix equal to,
\begin{equation}
\det{M_{\chi^0}^{PSS}}=-\mu^2 M_1 M_2 + \frac{1}{2} v_uv_d\mu \left( 
M_1 \tilde g_u\tilde g_d + M_2 \tilde g'_u \tilde g'_d \right),
\label{detNeut2}
\end{equation}
which is analogous to Eq.~(\ref{detNeut}).
\item {\textit {\bf SUSY SU(5) with Split SUSY}}

Now, let us discuss how one can find the R-parity violating 
couplings in the context of the simplest UV completion 
of the MSSM, the minimal SUSY SU(5). In this context the relevant 
superpotential is given by
\begin{equation}
{\cal W}_{NR}^{SU(5)}= 
\eta_i \hat{\bar{5}}_i \hat{5}_H \ + 
\ c_i \hat{\bar{5}}_i \hat{24}_H \hat{5}_H 
\ + \ \Lambda_{ijk} \hat{10}_i \hat{\bar{5}}_j \hat{\bar{5}}_k, 
\end{equation}
where our notation is $\bar{5}^T\!=\!(\hat{D}^C,-\hat{L}^T i \sigma_2)$\,, 
$10\!=\!(\hat{Q},\hat{U}^C,\hat{E}^C)$\,, 
$5_H^T\!=\!(\hat{T},\hat{H}_u)$\,, and 
$\hat{24}_H\!=\!(\hat{\Sigma}_8,\hat{\Sigma}_3,\hat{\Sigma}_{(3,2)}, 
\hat{\Sigma}_{(\bar{3},2)},\hat{\Sigma}_{24})$\,. 
Since all trilinear terms are coming from 
the same term in SU(5) one finds
\begin{equation}
\alpha_{ijk}/2=\beta_{ijk}=\gamma_{ijk}=\Lambda_{ijk}=-\Lambda_{ikj},
\end{equation} 
and the relevant interactions for the generation of neutrinos masses are 
given by
\begin{equation}
{\cal L}_{RpV}= - a_i \ \nu_i \ \tilde{H}_u^0 \ + 
\ \frac{1}{2} c_i \ \nu_i \ \tilde{\Sigma}_3^0 \ H_u^0 
\ + \ \frac{3 c_i}{ 2 \sqrt{15}} \nu_i \ \tilde{\Sigma}_{24} \ H_u^0 \ + 
\ \text{h.c.},
\end{equation}
where at the renormalizable level 
$M_{\Sigma_3}\!=\!5 M_{\Sigma_{24}}\!=\!M_{\Sigma}$. 
Therefore, in this case one has the usual contribution from the bilinear 
term plus 
an extra contribution for the neutrino masses once we integrate out the 
neutral 
component of $\Sigma_3$ and $\Sigma_{24}$. It is important to mention that
$a_i= \eta_i - 3 \langle\Sigma_{24}\rangle c_i / 2 \sqrt{15}$.  
Now, integrating out the 
fields $\Sigma_3$ and $\Sigma_{24}$ one finds that the mass matrix for 
neutrinos 
is given by
\begin{equation}
M^{SU(5)}_{ij} = M^{SS}_{ij} + \frac{v_u^2}{M_{\Sigma}}c_i c_j ,
\label{su5numass}
\end{equation}
where one can have $M_{\Sigma}\approx 10^{15-16}$ GeV in agreement with 
the unification constraints~\cite{German}. 
\end{itemize}
\section{ONE-LOOP CORRECTIONS TO THE NEUTRINO MASS MATRIX}
The one loop corrections are crucial for the correct characterization 
of neutrino phenomena. In the MSSM usually the most important one-loop 
contributions to the neutrino mass matrix are the bottom squarks, 
charginos, and neutralinos contributions. 
\subsection{Split SUSY Case}
In Split SUSY all scalars, except for one light Higgs boson, are superheavy. 
Therefore, in this case the only potentially important contributions 
are charginos and neutralinos together with $W$, $Z$, and light Higgs 
inside the loop. We show in Appendix A that $Z$ and $W$ loops are just 
a small renormalization of the tree-level contribution. The Higgs boson 
loop together with neutralinos has the same property in the decoupling 
limit. We discuss those contributions in detail in this section.

In general, the one loop contributions to the neutrino mass matrix 
can be written as~\cite{Hirsch:2000ef}:
\begin{equation}
\Delta M_{\nu}^{ij}=\Pi_{ij}(0)=-\frac{1}{16\pi^2}\sum_{f,b}
G_{ijfb}m_fB_0(0;m_f^2,m_b^2),
\label{DMgeneral}
\end{equation}
where the sum is over the fermions ($f$) and the bosons ($b$) inside the 
loop, $m_f$ is the fermion mass, and $G_{ijfb}$ is defined by the couplings
between the neutrinos and the fermions and bosons inside the loop. Once
the smallness of the $\epsilon_i$ and $\lambda_i$ parameters is taken
into account, each contribution can be expressed in the form
\begin{equation}
\Delta\Pi_{ij}=A^{(1)}\lambda_i\lambda_j+
B^{(1)}(\epsilon_i\lambda_j+\epsilon_j\lambda_i)+
C^{(1)}\epsilon_i\epsilon_j,
\label{deltapi}
\end{equation}
with $A^{(1)}$, $B^{(1)}$, and $C^{(1)}$ parameters independent of 
$\epsilon_i$ and $\lambda_i$, but dependent on the other SUSY parameters.
The super-index $(1)$ refers to the one-loop contribution. The tree-level
neutrino mass matrix in Eq.~(\ref{treenumass}) has the form
${\bf M}_{\nu\,\,\,ij}^{eff}=A^{(0)}\lambda_i\lambda_j$ with
\begin{equation}
A^{(0)}=\frac{v^2}{4\det{M_{\chi^0}^{SS}}}
\left(M_1 \tilde g^2_d + M_2 \tilde g'^2_d  \right),
\label{Atree}
\end{equation}
and we define the one-loop corrected parameters $A=A^{(0)}+A^{(1)}$,
$B=B^{(1)}$, and $C=C^{(1)}$.

In the MSSM with BRpV the neutral Higgs bosons mix with the sneutrinos 
forming two sets of 5 scalars and 5 pseudo-scalars. Nevertheless, in 
Split SUSY, all the sneutrinos are extremely heavy and decouple 
from the light Higgs boson $H$. In addition, the heavy Higgs boson 
also has a very large mass, leaving the light Higgs as the only neutral 
scalar able to contribute to the neutrino masses. This contribution is 
represented by the following Feynman graph, 
\begin{center}
\vspace{-50pt} \hfill \\
\begin{picture}(200,120)(0,23) 
%
%
\ArrowLine(20,50)(80,50)
\Text(50,60)[]{$\nu_j$}
\ArrowArcn(110,50)(30,180,0)
\Text(110,13)[]{$H$}
\DashCArc(110,50)(30,180,0){3}
\Text(110,95)[]{$\chi^0_k$}
\ArrowLine(140,50)(200,50)
\Text(170,60)[]{$\nu_i$}
\end{picture}
\vspace{30pt} \hfill \\
\end{center}
\vspace{-10pt}
which is proportional to the neutralino mass $m_{\chi^0_k}$. Here 
$\chi^0_k$ and $H$ are the neutralino and Higgs mass eigenstates,
but the graph is calculated in the basis where $\nu_i$ are not mass
eigenstates. The fields $\nu_i$ are the neutrino fields associated to 
the effective mass matrix given in Eq.~(\ref{treenumass}). This 
contribution to Eq.~(\ref{DMgeneral}) proceeds with the coupling 
\cite{Hirsch:2000ef}
\begin{equation}
G^h_{ijk}=\frac{1}{2}(O_{Ljk}^{nnh}O_{Lki}^{nnh}+O_{Rjk}^{nnh}O_{Rki}^{nnh}),
\label{Ghnn}
\end{equation}
where the relevant vertex is:
\begin{center}
\vspace{-50pt} \hfill \\
\begin{picture}(110,90)(0,23) 
\DashLine(10,25)(50,25){3}
\ArrowLine(50,25)(78,53)
\ArrowLine(78,-3)(50,25)
\Text(10,35)[]{$H$}
\Text(90,55)[]{$F_i^0$}
\Text(90,-5)[]{$F_j^0$}
\end{picture}
$
=\,i\,\Big[O^{nnh}_{Lij}\frac{(1-\gamma_5)}{2}+
O^{nnh}_{Rij}\frac{(1+\gamma_5)}{2}\Big]
$
\vspace{30pt} \hfill \\
\end{center}
\vspace{10pt}
Here $F^0_i$ are the seven eigenvectors linear combination of the 
higgsinos, gauginos, and neutrinos. The $O_L$ and $O_R$ couplings satisfy
$O^{nnh}_{Lij}=(O^{nnh}_{Rji})^*$ and above the scale $\tilde m$ we have,
\begin{eqnarray}
O^{nnh}_{Rij}&=&\half\Big\{
{\cal N}_{i4}\left(g s_\beta{\cal N}_{j2}-
g' s_\beta{\cal N}_{j1}\right)-
{\cal N}_{i3}\left(g c_\beta{\cal N}_{j2}-
g' c_\beta{\cal N}_{j1}\right)+
\nonumber\\
&&\qquad
{\cal N}_{i\,\ell+4}\left(g s_\ell{\cal N}_{j2}-
g' s_\ell{\cal N}_{j1}\right)+(i\leftrightarrow j)
\Big\},
\end{eqnarray}
where we have an implicit sum over $\ell=1,2,3$. We allow the matrix elements
of the matrix $\cal N$ to be imaginary when one of the eigenvalues is
negative, such that we do not need to include explicitly the sign called 
$\eta_i$ in Ref.~\cite{Hirsch:2000ef}. The difference with the 
MSSM couplings given in Ref.~\cite{Gunion:1989we} lies in the fact 
that in our case ${\cal N}$ is a $7\times7$ matrix, and the Higgs mixing angle 
has been replaced by $\alpha=\beta-\pi/2$, valid in the decoupling 
limit~\cite{Gunion:2002zf}. In addition, the third term is not present 
in the MSSM and comes from the second term in the supersymmetric 
lagrangian of Eq.~(9). 

Comparing the lagrangian below the scale $\widetilde m$ in 
Eq.~(\ref{LSplitRpV}) with the relevant term of the supersymmetric
lagrangian above $\widetilde m$ given by
\begin{equation}
{\cal L}_{SUSY} \owns 
-\frac{1}{\sqrt{2}}\widetilde L_i^\dagger \left(
g\sigma^a\widetilde W^a - g'\widetilde B\right) L_i,
\label{susyRpV}
\end{equation}
and considering the mixing between sleptons and Higgs bosons above that 
scale, a correspondence is found when the replacement
$\widetilde L^*_i\rightarrow-s_i i \sigma_2 H $ is made. The relevant 
matching condition at $\widetilde m$ is
\begin{equation}
a_i(\widetilde m)=\frac{s_i(\widetilde m)}{\cos\beta(\widetilde m)},
\label{matching}
\end{equation}
where the parameters $s_i(\widetilde m)$ represent the amount of slepton
$\widetilde L_i$ in the low energy Higgs $H$, and related to the sneutrino 
vev present above the scale $\tilde m$,
\begin{equation}
\tilde L_i=\left(\begin{matrix} \frac{1}{\sqrt{2}} 
(v_i+\tilde\ell^0_{si}+i\tilde\ell^0_{pi})
\cr \tilde\ell_{Li}^- \end{matrix}\right),
\end{equation}
as explained in the appendix. Using the approximation for the matrix 
${\cal N}$ from Appendix A, we obtain
for the coupling below the scale $\tilde m$:
\begin{eqnarray}
O_{Rik}^{\nu\chi h}&=&{\textstyle{1\over2}}\Big\{
-\left(\tilde g s_\beta N_{k2}-\tilde g' s'_\beta N_{k1}\right)\xi_{i4}-
N_{k4}\left(\tilde g s_\beta\,\xi_{i2}-\tilde g' s'_\beta\,\xi_{i1}\right)
\nonumber\\
&&\quad\;\,
+\left(\tilde g c_\beta N_{k2}-\tilde g' c'_\beta N_{k1}\right)
\left(\xi_{i3}-a_i\right)+
N_{k3}\left(\tilde g c_\beta\,\xi_{i2}-\tilde g' c'_\beta\,\xi_{i1}\right)
\Big\}.
\end{eqnarray}
Notice that there is no term proportional to $\epsilon_i$ since there is a 
cancellation in $\xi_{i3}-a_i$. It can be checked using Eqs.~(\ref{xideff}) 
and the definition of $\lambda_i=a_i\mu+\epsilon_i$. This implies that the 
contribution of the light Higgs boson has the form:
\begin{equation}
\Delta\Pi_{ij}^{h}=A^h\lambda_i\lambda_j,
\label{DpiH}
\end{equation}
which does not break the symmetry of the neutrino mass matrix at tree level.
The detailed expression is given by,
\begin{eqnarray}
\Delta\Pi_{ij}^h&=&-\frac{1}{16\pi^2}\sum_{k=1}^4
(\widetilde O^{\nu\chi h}_k)^2\lambda_i\lambda_j
m_{\chi_k^0}B_0(0;m_{\chi_k^0}^2,m_h^2),
\label{Splitloop}
\end{eqnarray}
with
\begin{eqnarray}
\widetilde O_k^{\nu\chi h}&=&{\textstyle{1\over2}}\Big\{
-\left(\tilde g s_\beta N_{k2}-\tilde g' s'_\beta N_{k1}\right)\xi_4-
N_{k4}\left(\tilde g s_\beta\,\xi_2-\tilde g' s'_\beta\,\xi_1\right)
\nonumber\\
&&\quad\;\,
+\left(\tilde g c_\beta N_{k2}-\tilde g' c'_\beta N_{k1}\right)
\left(\xi_3-1/\mu\right)+
N_{k3}\left(\tilde g c_\beta\,\xi_2-\tilde g' c'_\beta\,\xi_1\right)
\Big\}.
\end{eqnarray}
Since the gauge and Goldstone boson contribute to the neutrino mass matrix 
in the same form, as can be checked in the Appendix, we conclude that it 
is not possible to generate the neutrino masses in Split Supersymmetry with
bilinear R-Parity violating interactions alone. This conclusion is in 
agreement with the results presented in Ref.~\cite{Chun:2004mu}, and 
in Ref.~\cite{Davidson:2000ne}, where the contribution from the Higgs 
boson can be inferred taking the decoupling limit (see also
\cite{Davidson:2000uc}).
\subsection{Partial Split SUSY Case}
In this scenario the five physical Higgs states, $h,H,A,H^\pm$, are light
and contribute to the neutrino mass matrix. In the following subsections
we divide them in CP-even, CP-odd, and charged Higgs contributions.
\subsubsection{\bf CP-even neutral Higgs bosons}
The two CP-even neutral Higgs bosons contribute to
the neutrino mass matrix through the following graphs,
\begin{center}
\vspace{-50pt} \hfill \\
\begin{picture}(200,120)(0,23) 
%
%
\ArrowLine(20,50)(80,50)
\Text(50,60)[]{$\nu_j$}
\ArrowArcn(110,50)(30,180,0)
\Text(110,13)[]{$h,H$}
\DashCArc(110,50)(30,180,0){3}
\Text(110,95)[]{$\chi^0_k$}
\ArrowLine(140,50)(200,50)
\Text(170,60)[]{$\nu_i$}
\end{picture}
\vspace{30pt} \hfill \\
\end{center}
\vspace{-10pt}
where the $G$ factor in Eq.~(\ref{DMgeneral}) is,
\begin{equation}
G^s_{ijkr}=\frac{1}{2}(O_{Ljkr}^{nns}O_{Lkir}^{nns}+O_{Rjkr}^{nns}O_{Rkir}^{nns}).
\label{Gsnn}
\end{equation}
The relevant coupling above the scale $\tilde m$ is the CP-even neutral
scalar couplings to two neutral fermions, given by,
\begin{center}
\vspace{-50pt} \hfill \\
\begin{picture}(110,90)(0,23) 
\DashLine(10,25)(50,25){3}
\ArrowLine(50,25)(78,53)
\ArrowLine(78,-3)(50,25)
\Text(10,35)[]{$S^0_k$}
\Text(90,55)[]{$F_j^0$}
\Text(90,-5)[]{$F_i^0$}
\end{picture}
$
=\,i\,\Big[O^{nns}_{Lijk}\frac{(1-\gamma_5)}{2}+
O^{nns}_{Rijk}\frac{(1+\gamma_5)}{2}\Big]
$
\vspace{30pt} \hfill \\
\end{center}
\vspace{10pt}
where,
\begin{eqnarray}
O_{Lijk}^{nns}&=&\half\left[
\left(-R^0_{k1}{\cal N}^*_{j3}+R^0_{k2}{\cal N}^*_{j4}
-R^0_{k\,\ell+2}{\cal N}^*_{j\,\ell+4}\right)
\left(g{\cal N}^*_{i2}-g'{\cal N}^*_{i1}\right)
+(i\leftrightarrow j)\right],
\label{Onns}
\end{eqnarray}
and $O_{Rijk}^{nns}=(O_{Lijk}^{nns})^*$.
The fields $S_k^0$ are linear combinations of CP-even Higgs and 
sneutrinos whose mass matrix in the basis 
$(\phi^0_d , \phi^0_u , \tilde\ell^0_{si})$ is given in the Appendix B.
In the PSSusy, the mass matrix can be diagonalized by,
\begin{equation}
\left(\begin{matrix}
h \cr H \cr \tilde\nu^i_s
\end{matrix}\right)=
\left(\begin{matrix}
-s_\alpha & c_\alpha & -s_s^j \cr
c_\alpha & s_\alpha & -t_s^j \cr
-s_\alpha s_s^i+c_\alpha t_s^i & c_\alpha s_s^i+s_\alpha t_s^i & \delta_{ij}
\end{matrix}\right)
\left(\begin{matrix}
\phi^0_d \cr \phi^0_u \cr \tilde\ell^0_{sj}
\end{matrix}\right),
\end{equation}
where the angle $\alpha$ is analogous to the CP-even neutral Higgs bosons
mixing angle of he MSSM. 
An expression for the mixing angles $s_s^i$ and $t_s^i$ above the
scale $\tilde m$ can be found in the Appendix B. Comparing the 
supersymmetric lagrangian above he scale $\tilde m$ in Eq.~(\ref{susyRpV})
with the terms of the PSSusy lagrangian in Eq.~(\ref{LSS2HDMRpV}) we
find the following matching conditions,
\begin{equation}
s_s^i(\tilde m)=-b_i(\tilde m) c_\alpha
\,;\qquad
t_s^i(\tilde m)=-b_i(\tilde m) s_\alpha,
\label{st_s}
\end{equation}
where $s_s^i(\tilde m)$ represents the amount of slepton $\tilde L_i$
present in the low energy light Higgs $h$, and analogously with 
$t_s^i(\tilde m)$ for the low energy heavy Higgs $H$. In the limit where 
the sleptonic fields have a very large mass, they satisfy,
\begin{equation}
s_s^i\rightarrow-c_\alpha\frac{v_i}{v_u}\,,\qquad
t_s^i\rightarrow-s_\alpha\frac{v_i}{v_u},
\end{equation}
which tells us that the parameter $b_i$, defined below $\tilde m$, is 
directly proportional to the sneutrino vacuum expectation value $v_i$, 
defined above the scale $\tilde m$.

In the coupling in Eq.~(\ref{Onns}), we take the first neutral fermion as 
a neutrino and the second as a neutralino, obtaining the following 
couplings for both Higgs bosons $h$ and $H$,
\begin{eqnarray}
O_{Lik}^{\nu\chi h}&=&\half\bigg[
\left(s_\alpha N^*_{k3}+c_\alpha N^*_{k4}\right)
\left(-g\xi_{i2}+g'\xi_{i1}\right)
+
\left(-s_\alpha\xi_{i3}-c_\alpha\xi_{i4}+s_s^i\right)
\left(g N^*_{k2}-g' N^*_{k1}\right)
\bigg],
\nonumber\\
O_{Lik}^{\nu\chi H}&=&\half\bigg[
\left(-c_\alpha N^*_{k3}+s_\alpha N^*_{k4}\right)
\left(-g\xi_{i2}+g'\xi_{i1}\right)
+
\left(c_\alpha\xi_{i3}-s_\alpha\xi_{i4}+t_s^i\right)
\left(g N^*_{k2}-g' N^*_{k1}\right)
\bigg].
\end{eqnarray}
After isolating the terms proportional to $\epsilon_i$ 
in the couplings, and using eq.~(\ref{st_s}), we find the following 
expressions valid below $\tilde m$,
\begin{eqnarray}
O_{Lik}^{\nu\chi h}&=&\widetilde O_{Lk}^{\nu\chi h}\Lambda_i+
\frac{1}{2\mu s_\beta}
\cos(\alpha-\beta)
\left(g N^*_{k2}-g' N^*_{k1}\right)\epsilon_i,
\nonumber\\
O_{Lik}^{\nu\chi H}&=&\widetilde O_{Lk}^{\nu\chi H}\Lambda_i+
\frac{1}{2\mu s_\beta}
\sin(\alpha-\beta)
\left(g N^*_{k2}-g' N^*_{k1}\right)\epsilon_i,
\label{OnuXh}
\end{eqnarray}
with the term proportional to $\Lambda_i$ given by,
\begin{eqnarray}
\widetilde O_{Lk}^{\nu\chi h}&=&-\frac{1}{2}\bigg[
\left(s_\alpha N^*_{k3}+c_\alpha N^*_{k4}\right)
\left(g\xi_2-g'\xi_1\right)
+
\left(s_\alpha\xi_3+c_\alpha\xi_4
+\frac{c_\alpha}{\mu v_u} \right)
\left(g N^*_{k2}-g' N^*_{k1}\right)
\bigg],
\nonumber\\
\widetilde O_{Lk}^{\nu\chi H}&=&\frac{1}{2}\bigg[
\left(c_\alpha N^*_{k3}-s_\alpha N^*_{k4}\right)
\left(g\xi_2-g'\xi_1\right)
+
\left(c_\alpha\xi_3-s_\alpha\xi_4
-\frac{s_\alpha}{\mu v_u} \right)
\left(g N^*_{k2}-g' N^*_{k1}\right)
\bigg].
\end{eqnarray}
Notice that the presence of the term proportional to $\epsilon_i$ 
in Eq.~(\ref{OnuXh}) implies that the contribution of the CP-even 
Higgs bosons has the form:
\begin{equation}
\Delta\Pi_{ij}=A\Lambda_i\Lambda_j+
B(\Lambda_i\epsilon_j+\Lambda_j\epsilon_i)+
C\epsilon_i\epsilon_j,
\label{DpiH2HDM}
\end{equation}
breaking the symmetry of the neutrino mass matrix at tree level, and 
generating a solar mass. Explicitly, this contribution is:
\begin{eqnarray}
\Delta\Pi_{ij}^{hH}&=&-\frac{1}{16\pi^2}\sum_{k=1}^4\sum_{n=1}^2
\left(E_k^n\Lambda_i+F_k^n\epsilon_i\right)
\left(E_k^n\Lambda_j+F_k^n\epsilon_j\right)
m_{\chi_k^0}B_0(0;m_{\chi_k^0}^2,m_{H_n}^2),
\label{2loops}
\end{eqnarray}
with
\begin{eqnarray}
& E_k^1 = \widetilde O_{Lk}^{\nu\chi h} \,,\qquad
& F_k^1 = \frac{\cos(\alpha-\beta)}{2\mu s_\beta}
\left(g N^*_{k2}-g' N^*_{k1}\right),
\nonumber\\
& E_k^2 = \widetilde O_{Lk}^{\nu\chi H} \,,\qquad
& F_k^2 =\frac{\sin(\alpha-\beta)}{2\mu s_\beta}
\left(g N^*_{k2}-g' N^*_{k1}\right),
\label{EandF}
\end{eqnarray}
where we work in the Feynman gauge. 
\subsubsection{\bf CP-odd neutral Higgs bosons}
Loops including the CP-odd Higgs boson $A$ must be added through the 
graph, 
\begin{center}
\vspace{-50pt} \hfill \\
\begin{picture}(200,120)(0,23) 
%
%
\ArrowLine(20,50)(80,50)
\Text(50,60)[]{$\nu_j$}
\ArrowArcn(110,50)(30,180,0)
\Text(110,13)[]{$A$}
\DashCArc(110,50)(30,180,0){3}
\Text(110,95)[]{$\chi^0_k$}
\ArrowLine(140,50)(200,50)
\Text(170,60)[]{$\nu_i$}
\end{picture}
\vspace{30pt} \hfill \\
\end{center}
\vspace{-10pt}
where the $G$ factor in Eq.~(\ref{DMgeneral}) is,
\begin{equation}
G^p_{ijkr}=-\frac{1}{2}(O_{Ljkr}^{nnp}O_{Lkir}^{nnp}+O_{Rjkr}^{nnp}O_{Rkir}^{nnp}).
\label{Gpnn}
\end{equation}
The relevant coupling above the scale $\tilde m$ is the CP-odd neutral 
scalar couplings to two neutral fermions, given by,
\begin{center}
\vspace{-50pt} \hfill \\
\begin{picture}(110,90)(0,23) 
\DashLine(10,25)(50,25){3}
\ArrowLine(50,25)(78,53)
\ArrowLine(78,-3)(50,25)
\Text(10,35)[]{$P^0_k$}
\Text(90,55)[]{$F_j^0$}
\Text(90,-5)[]{$F_i^0$}
\end{picture}
$
=\,\Big[O^{nnp}_{Lijk}\frac{(1-\gamma_5)}{2}+
O^{nnp}_{Rijk}\frac{(1+\gamma_5)}{2}\Big]
$
\vspace{30pt} \hfill \\
\end{center}
\vspace{10pt}
where,
\begin{eqnarray}
O_{Lijk}^{nnp}&=&-\half\left[
\left(-R^p_{k1}{\cal N}^*_{j3}+R^p_{k2}{\cal N}^*_{j4}
-R^p_{k\,\ell+2}{\cal N}^*_{j\,\ell+4}\right)
\left(g{\cal N}^*_{i2}-g'{\cal N}^*_{i1}\right)
+(i\leftrightarrow j)\right],
\label{Onnp}
\end{eqnarray}
and $O_{Rijk}^{nnp}=-(O_{Ljik}^{nnp})^*$.
The fields $P_k^0$ are linear combinations of CP-odd Higgs and 
sneutrinos whose mass matrix in the basis 
$(\varphi^0_d , \varphi^0_u , \tilde\ell^0_{pi})$ is given in the Appendix B.
In the PSSusy, the mass matrix can be diagonalized by,
\begin{equation}
\left(\begin{matrix}
G \cr A \cr \tilde\nu^i_p
\end{matrix}\right)=
\left(\begin{matrix}
-c_\beta & s_\beta & -s_p^j \cr
 s_\beta & c_\beta & -t_p^j \cr
-c_\beta s_p^i+s_\beta t_p^i & s_\beta s_p^i+c_\beta t_p^i & \delta_{ij}
\end{matrix}\right)
\left(\begin{matrix}
\varphi^0_d \cr \varphi^0_u \cr \tilde\ell^0_{pj}
\end{matrix}\right).
\end{equation}
An expression for the mixing angles $s_p^i$ and $t_p^i$ above the
scale $\tilde m$ can be found in the Appendix B. Comparing the 
supersymmetric lagrangian above he scale $\tilde m$ in Eq.~(\ref{susyRpV})
with the terms of the PSSusy lagrangian in Eq.~(\ref{LSS2HDMRpV}) we
find the following matching conditions,
\begin{equation}
s_p^i(\tilde m)=b_i(\tilde m) s_\beta
\,;\qquad
t_p^i(\tilde m)=b_i(\tilde m) c_\beta,
\label{st_p}
\end{equation}
where $s_p^i(\tilde m)$ represents the amount of slepton $\tilde L_i$
present in the Goldstone boson $G$, and analogously with 
$t_p^i(\tilde m)$ for the low energy CP-odd Higgs $A$. In the limit where 
the sleptonic fields have a very large mass,
\begin{equation}
s_p^i\rightarrow s_\beta\frac{v_i}{v_u}\,,\qquad
t_p^i\rightarrow c_\beta\frac{v_i}{v_u},
\end{equation}
which indicates $b_i = v_i/v_u$ in agreement with the CP-even 
case.

If we take the coupling in Eq.~(\ref{Onnp}) and expand on small
R-Parity violating parameters we find for the CP-odd Higgs bosons
couplings,
\begin{equation}
O_{Lik}^{\nu\chi a}=-\half\bigg[
\left(-s_\beta N^*_{k3}+c_\beta N^*_{k4}\right)
\left(-g\xi_{i2}+g'\xi_{i1}\right)
+
\left(s_\beta\xi_{i3}-c_\beta\xi_{i4}+t_p^i\right)
\left(g N^*_{k2}-g' N^*_{k1}\right)
\bigg].
\end{equation}
If we isolate the terms proportional to $\epsilon_i$, using 
eq.~(\ref{st_p}), we find,
\begin{equation}
O_{Lik}^{\nu\chi a}=\widetilde O_{Lk}^{\nu\chi a}\Lambda_i+
\frac{1}{2\mu s_\beta}
\left(g N^*_{k2}-g' N^*_{k1}\right)\epsilon_i.
\label{OnuXh2}
\end{equation}
It is shown in Appendix A that the Goldstone boson contribution completely cancels out
when gauge dependent terms from gauge couplings and tadpoles are included.
The $\widetilde O$ coupling is defined by,
\begin{equation}
\widetilde O_{Lk}^{\nu\chi a}=-\frac{1}{2}\bigg[
\left(s_\beta N^*_{k3}-c_\beta N^*_{k4}\right)
\left(g\xi_2-g'\xi_1\right)
+
\left(s_\beta\xi_3-c_\beta\xi_4
+\frac{c_\beta}{\mu v_u} \right)
\left(g N^*_{k2}-g' N^*_{k1}\right)
\bigg].
\end{equation}
In this way, the CP-odd contribution is
\begin{eqnarray}
\Delta\Pi_{ij}^{A}&=&\frac{1}{16\pi^2}\sum_{k=1}^4
\left(E_k^3\Lambda_i+F_k^3\epsilon_i\right)
\left(E_k^3\Lambda_j+F_k^3\epsilon_j\right)
m_{\chi_k^0}B_0(0;m_{\chi_k^0}^2,m_A^2),
\label{3loops}
\end{eqnarray}
with
\begin{equation}
E_k^3 = \widetilde O_{Lk}^{\nu\chi a} \,,\qquad
F_k^3 = \frac{1}{2\mu s_\beta}
\left(g N^*_{k2}-g' N^*_{k1}\right).
\label{EandF2}
\end{equation}
Note that the CP-odd contribution in Eq.~(\ref{EandF2}) has the opposite sign 
of the CP-even contribution. In addition, the $\epsilon_i\epsilon_j$ terms
in the limit of equal neutral Higgs masses. This is because the CP-even terms 
are proportional to $\cos^2(\alpha-\beta)B_0(0;m_{\chi_k^0}^2,m_h^2)$ and 
$\sin^2(\alpha-\beta)B_0(0;m_{\chi_k^0}^2,m_H^2)$, while the CP-odd
term is proportional to $-B_0(0;m_{\chi_k^0}^2,m_A^2)$. 
\section{Numerical Results}
\subsection{Partial Split SUSY}
As seen in the previous chapters, Partial Split Supersymmetry is determined
by the following supersymmetric parameters: the supersymmetric Higgs mass 
$\mu$, the gaugino masses $M_1$ and $M_2$, the mass of the lightest CP-even 
Higgs $m_h$, the CP-odd Higgs mass $m_A$, and the tangent of the CP-odd 
Higgs mixing angle $\tan\beta$.
As a working scenario we choose the numerical values given in Table I. 
\begin{table}
\begin{center}
\caption{PSS and neutrino mass matrix parameters.}
\bigskip
\begin{minipage}[t]{0.8\textwidth}
\begin{tabular}{ccc}
\hline
Parameter & Solution & Units \\
\hline \hline
$\tan\beta$ & 10 & -  \\
$\mu$ & 450 & GeV  \\
$M_2$ & 300 & GeV  \\
$M_1$  & 150 & GeV  \\
$m_h$  & 120 & GeV  \\
$m_A$  & 1000 & GeV  \\
$Q$  & 830 & GeV  \\
\hline 
$A$ & -2.7 & eV/GeV${}^4$ \\
$B$ & -0.0005 & eV/GeV${}^3$ \\
$C$ & 0.315 & eV/GeV${}^2$ \\
\hline \hline \label{tab:MSSMabc}
\end{tabular}
\end{minipage}
\end{center}
\end{table}
In this scenario the four neutralino masses are $m_{\chi}= 147, 282, 455, 476$
GeV, with the lightest neutralino the LSP. In the Higgs sector, the charged
Higgs mass is $m_{H^+}=1003.2$ GeV, the heavy neutral CP-even Higgs mass is 
$m_H=1000.2$ GeV, and the CP-even Higgs mixing angle is given by 
$\sin\alpha=0.101$.

The one-loop corrected parameters $A$, $B$, and $C$ introduced in 
Eq.~(\ref{DpiH2HDM}) are calculated with the results in Eq.~(\ref{2loops})
for the neutral CP-even Higgs bosons, in Eq.~(\ref{3loops}) for the neutral
CP-odd Higgs boson, and in Eq.~(\ref{4loops}) for the charged Higgs boson.
These contributions give rise to a set of parameters $A$, $B$, and $C$
given in Table I. The value of $A=-2.7 \, {\mathrm{eV/GeV}}^2$ is mainly due
to the tree level contribution, and $C=0.315 \, {\mathrm{eV/GeV}}^4$ is
completely generated by radiative corrections. 

The parameter $C$ is subtraction scale independent, while the parameters 
$A$ and $B$ depend on the subtraction scale $Q$. As a way of fixing this 
scale, we have chosen $Q$ such that it minimizes the parameter $B$, making 
the solar mass completely scale independent. For the scenario in Table I
we find that $Q=830$ GeV gives rise to $B=-0.0005 \, {\mathrm{eV/GeV}}^3$, 
which is already negligible.

We notice that in the decoupling limit scenario the light CP-even Higgs 
$h$ contribution to the solar mass (or equivalently, to the parameter $C$) 
is negligible, since it is proportional to $\cos(\alpha-\beta)\rightarrow 0$.
Therefore, it can be said properly that the solar mass comes exclusively 
from the contributions of the heavy Higgs bosons $H$ and $A$. Further more, 
as indicated by Eqs.~(\ref{EandF}) and (\ref{EandF2}) the contributions 
from $H$ and $A$ have opposite signs and tend to cancel each other in the 
decoupling limit, where $\sin(\alpha-\beta)\rightarrow 1$ and 
$m_H\rightarrow m_A$. In our scenario, $\cos(\alpha-\beta)=0.0016$ and
$m_H-m_A=0.2$ GeV, and the cancellation between $H$ and $A$ contributions
to $C$ is at the $0.07\%$.
\begin{table}
\begin{center}
\caption{BRpV parameters and neutrino observables.}
\bigskip
\begin{minipage}[t]{0.8\textwidth}
\begin{tabular}{ccc}
\hline
Parameter & Solution & Units \\
\hline \hline
$\epsilon_1$ & 0.0346 & GeV  \\
$\epsilon_2$ & 0.265 & GeV  \\
$\epsilon_3$ & 0.322 & GeV  \\
$\Lambda_1$  & -0.0269 & GeV${}^2$  \\
$\Lambda_2$  & -0.00113 & GeV${}^2$  \\
$\Lambda_3$  & 0.0693 & GeV${}^2$  \\
\hline $\Delta m_{atm}^2$ & 2.34$\times10^{-3}$ & eV${}^2$ \\
$\Delta m_{sol}^2$ & 8.16$\times10^{-5}$ & eV${}^2$ \\
$\tan^2\theta_{atm}$ & 1.04 & - \\
$\tan^2\theta_{sol}$ & 0.455 & - \\
$\tan^2\theta_{13}$ & 0.0247 & - \\
$m_{ee}$ & 0.00394 & eV \\
\hline \hline \label{tab:BRpVsol}
\end{tabular}
\end{minipage}
\end{center}
\end{table}
Within the scenario in Table I we look for a solution to the neutrino 
observables varying $\vec\epsilon$ and $\vec\Lambda$. An example solution 
is given in Table II. This solution satisfy 
$\epsilon_1\ll\epsilon_2,\epsilon_3$ and 
$|\Lambda_2|\ll|\Lambda_1|,\Lambda_3$. The sign of these parameters have
a very small influence. Also in Table II we list the neutrino observables.
The atmospheric mass $\Delta m^2_{atm}=2.34\times10^{-3}\,{\mathrm{eV}}^2$
and the solar mass $\Delta m^2_{sol}=8.16\times10^{-5}\,{\mathrm{eV}}^2$
are practically at the center of the experimentally allowed regions. The
atmospheric angle $\tan^2\theta_{atm}=1.04$ is slightly deviated from
maximal mixing, while the solar angle $\tan^2\theta_{sol}=0.455$ is 
non-maximal and with a value centered on the experimentally allowed region.
The other two parameters, the reactor angle $\tan^2\theta_{13}=0.0247$ and
the neutrinoless double beta decay mass $m_{ee}=0.00394$ eV have not been 
experimentally measured and the predictions of our model are well below the 
experimental upper bounds.

\begin{figure}
\centerline{\protect\hbox{\epsfig{file=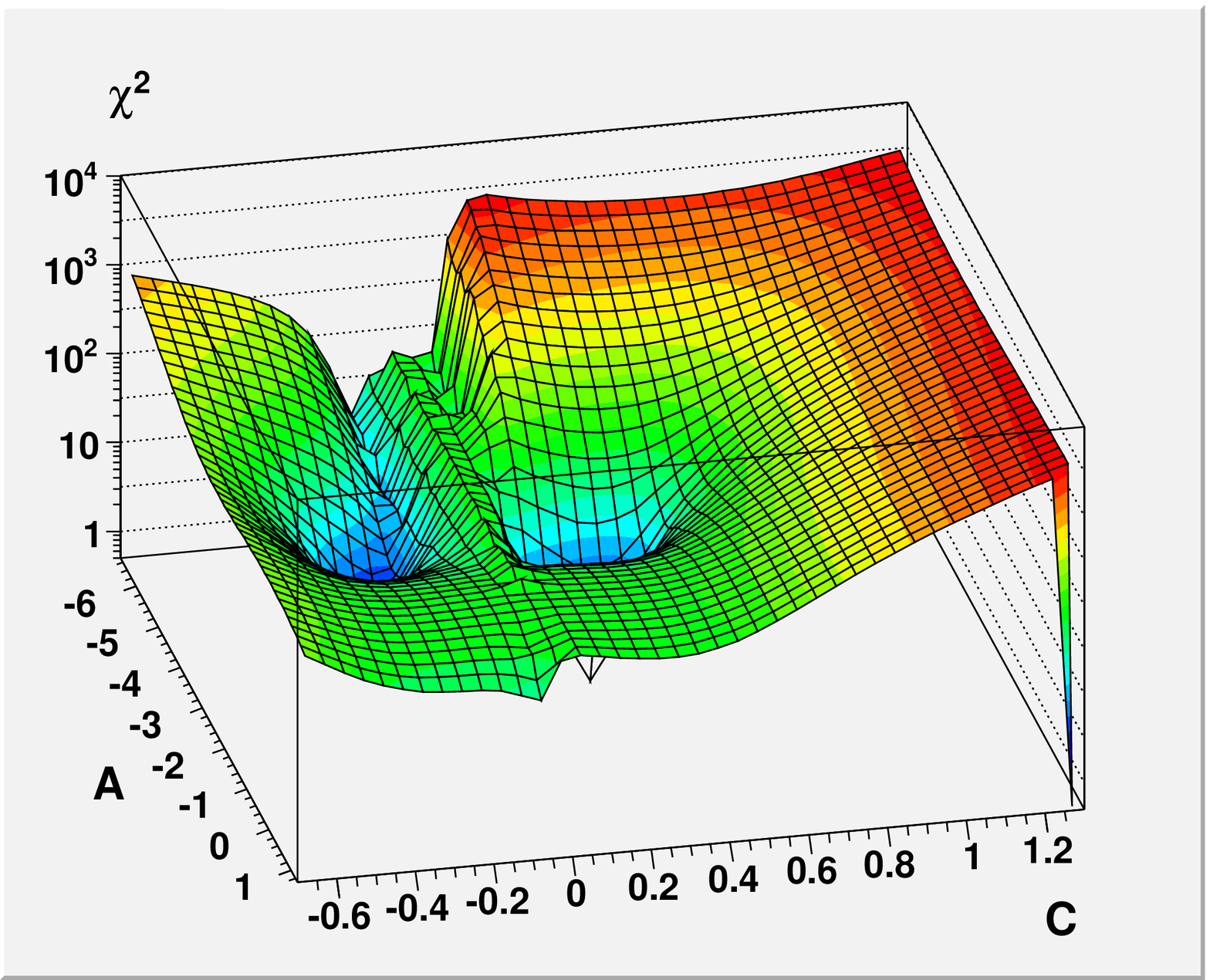,width=0.5\textwidth},
\epsfig{file=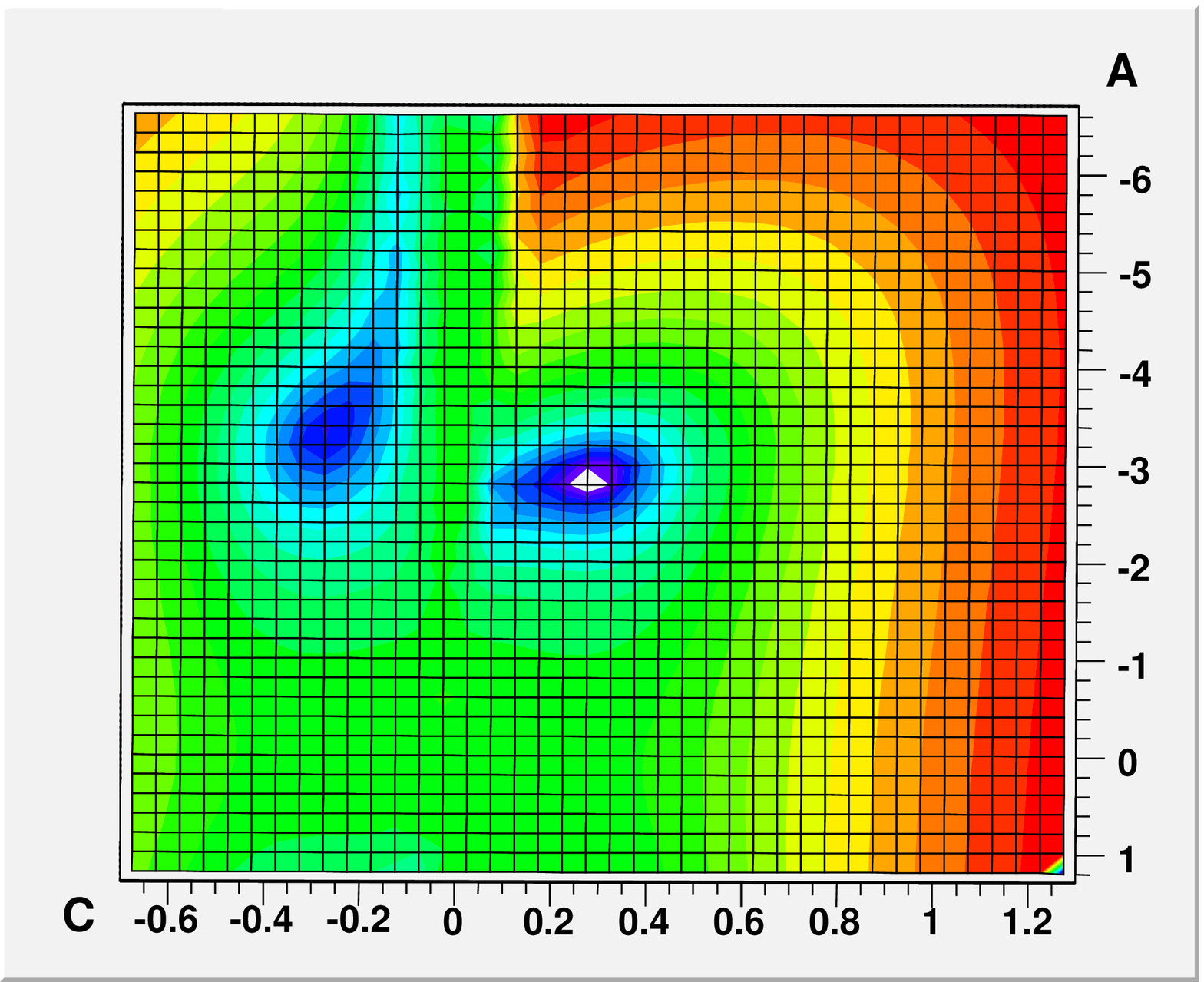,width=0.5\textwidth}}}
\caption{\it Neutrino physics $\chi^2$ as a function of the neutrino mass
matrix parameters $A$ and $C$, keeping $\vec\epsilon$ and $\vec\Lambda$
as indicated in Table I.
}
\label{AC}
\end{figure}
In order to study the dependence of the neutrino physics solutions on 
different parameters we have implemented the following $\chi^2$
\begin{equation}
\chi^2=\left(\frac{10^3\Delta m^2_{atm}-2.35}{0.95}\right)^2+
\left(\frac{10^5\Delta m^2_{sol}-8.15}{0.95}\right)^2+
\left(\frac{\sin^2\theta_{atm}-0.51}{0.17}\right)^2+
\left(\frac{\sin^2\theta_{sol}-0.305}{0.075}\right)^2.
\end{equation}
In each of these terms we evaluated how many standard deviation the prediction
is from the measured experimental central values \cite{Maltoni:2004ei}. 
In Fig.~\ref{AC} we have 
$\chi^2$ in the vertical axis as a function of $A$ and $C$, in perspective in
the left frame and level contours in the right frame. The preferred solution 
of Table I appears at the center of the graphs. Neutrino observables are very
sensitive to the parameters $A$ and $C$ as shown by contours, where the
darkest ellipsoid (blue) corresponds to $\chi^2\lsim10$, while the white 
center corresponds to $\chi^2\lsim1$. There is a second minima, but it does
not reach values near unity.

A good approximation for the neutrino masses in this scenario is the 
following,
\begin{eqnarray}
m_3&=&C|\vec\epsilon|^2
+A\frac{(\vec\epsilon\cdot\vec\Lambda)^2}{|\vec\epsilon|^2},
\nonumber\\
m_2&=&
A\frac{|\vec\epsilon\times(\vec\Lambda\times\vec\epsilon)|^2}
{|\vec\epsilon|^4},
\label{miapp}
\end{eqnarray}
with the third neutrino massless \cite{Diaz:2004fu}. Despite the fact 
that $C$ is one-loop
generated and $A$ receives contributions at tree level, the first term
in $m_3$ is dominant, and thus more important for the atmospheric mass
scale. The $A$ term is the only one contributing to the solar mass, as 
indicated in Eq.~(\ref{miapp}).

\begin{figure}
\centerline{\protect\hbox{\epsfig{file=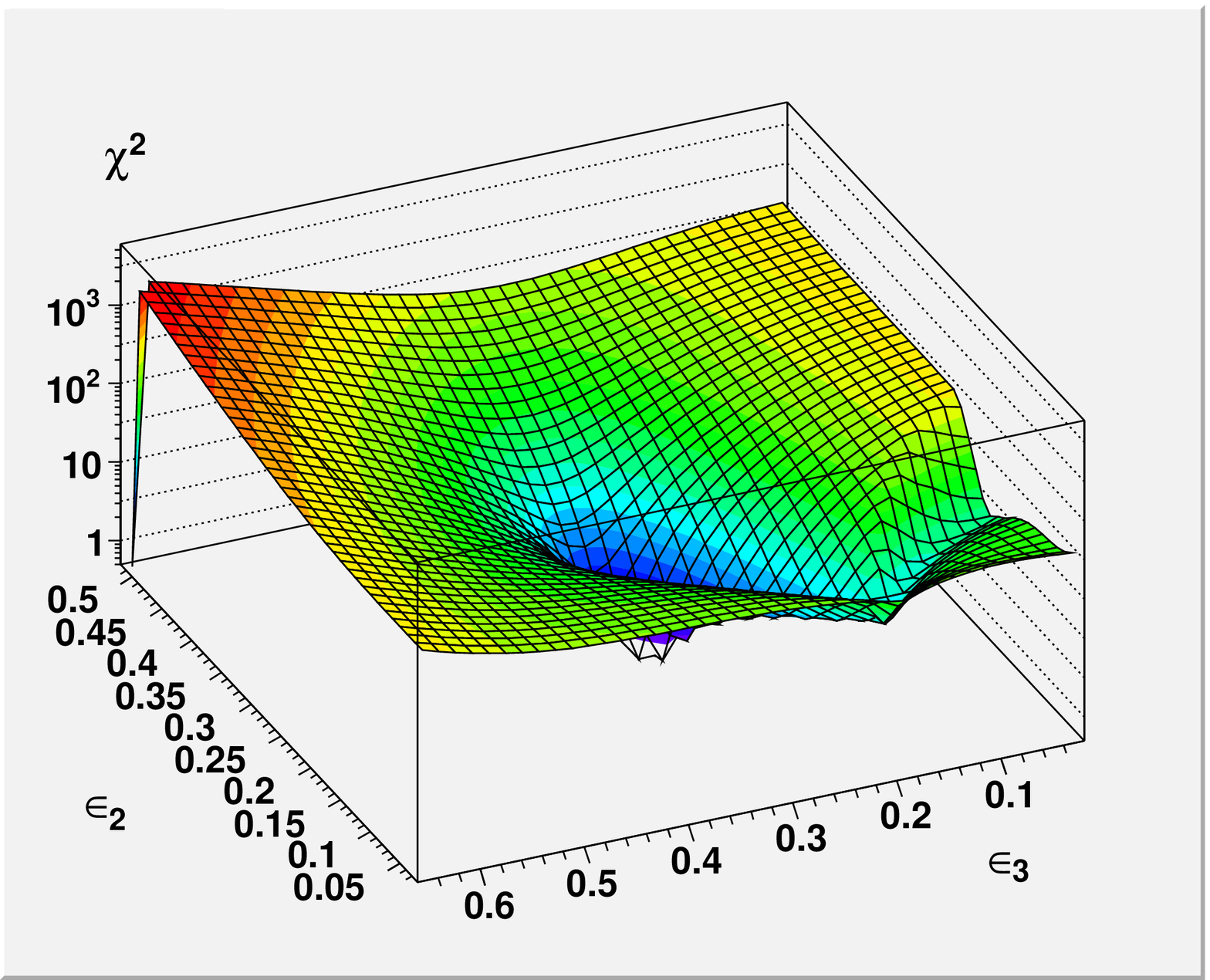,width=0.5\textwidth},
\epsfig{file=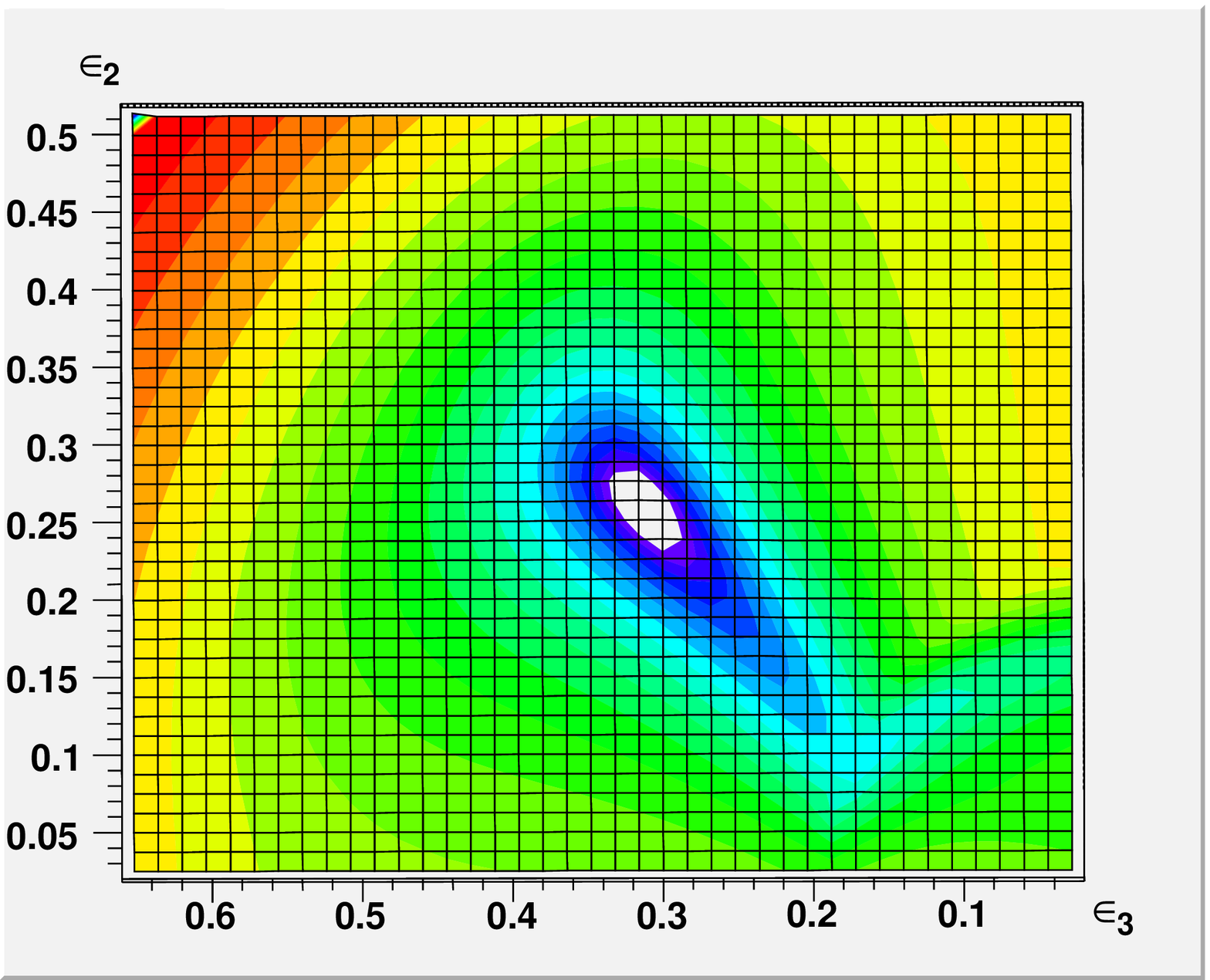,width=0.5\textwidth}}}
\caption{\it Neutrino physics $\chi^2$ as a function of the BRpV parameters 
$\epsilon_2$ and $\epsilon_3$, keeping the rest of the parameters as 
indicated in Table I.
}
\label{e2e3}
\end{figure}
In Fig.~\ref{e2e3} we plot $\chi^2$ as a function of $\epsilon_2$ and 
$\epsilon_3$ in two frames as described for the previous figure. The rest 
of the BRpV parameters are fixed to the values in Table I, while the 
values of $A$ and $C$ are calculated from the loop contributions.
In our scenario approximated expression can be found when $\epsilon_1$
and $\Lambda_2$ are neglected. It turns out that the atmospheric angle and
mass squared difference depend strongly on $\epsilon_2$ and $\epsilon_3$.
They are given by,
\begin{eqnarray}
\Delta m_{\rm atm}^2 &\approx& C^2(\epsilon_2^2+\epsilon_3^2)^2,
\nonumber\\
\tan^2\theta_{\rm atm} &\approx& 
\left(\frac{\epsilon_2}{\epsilon_3}\right)^2.
\label{approximatm}
\end{eqnarray}
Notice that it is the atmospheric mass who receives the main contribution 
from loop corrections, with $C$ generated entirely at one-loop. Equal values
for the atmospheric mass correspond to circles around the origen in the 
$\epsilon_2$-$\epsilon_3$ plane, while equal values for the atmospheric angle
are represented by straight lines passing through the origen. This geometry
can be visualized in Fig.~\ref{e2e3}.

\begin{figure}
\centerline{\protect\hbox{\epsfig{file=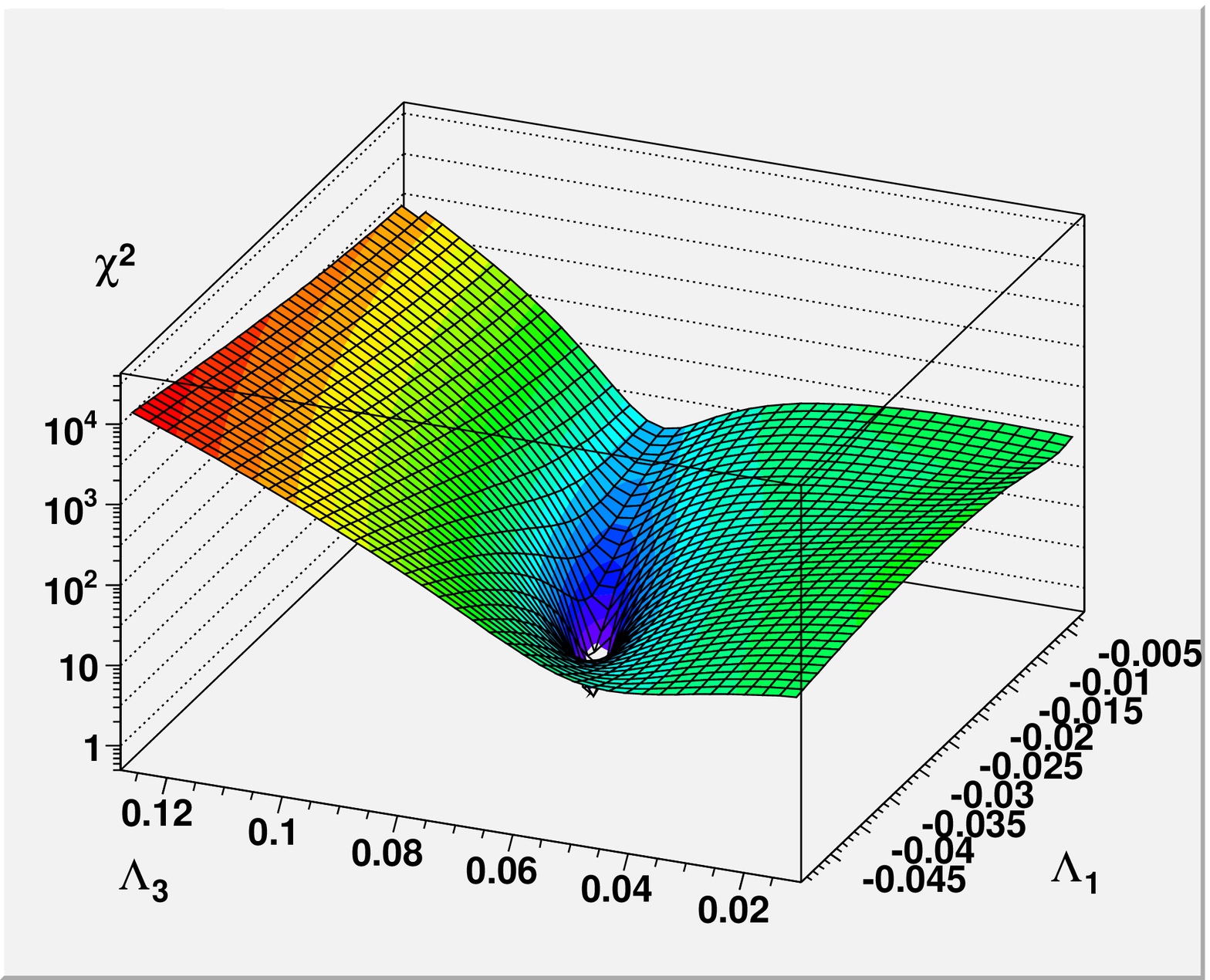,width=0.5\textwidth},
\epsfig{file=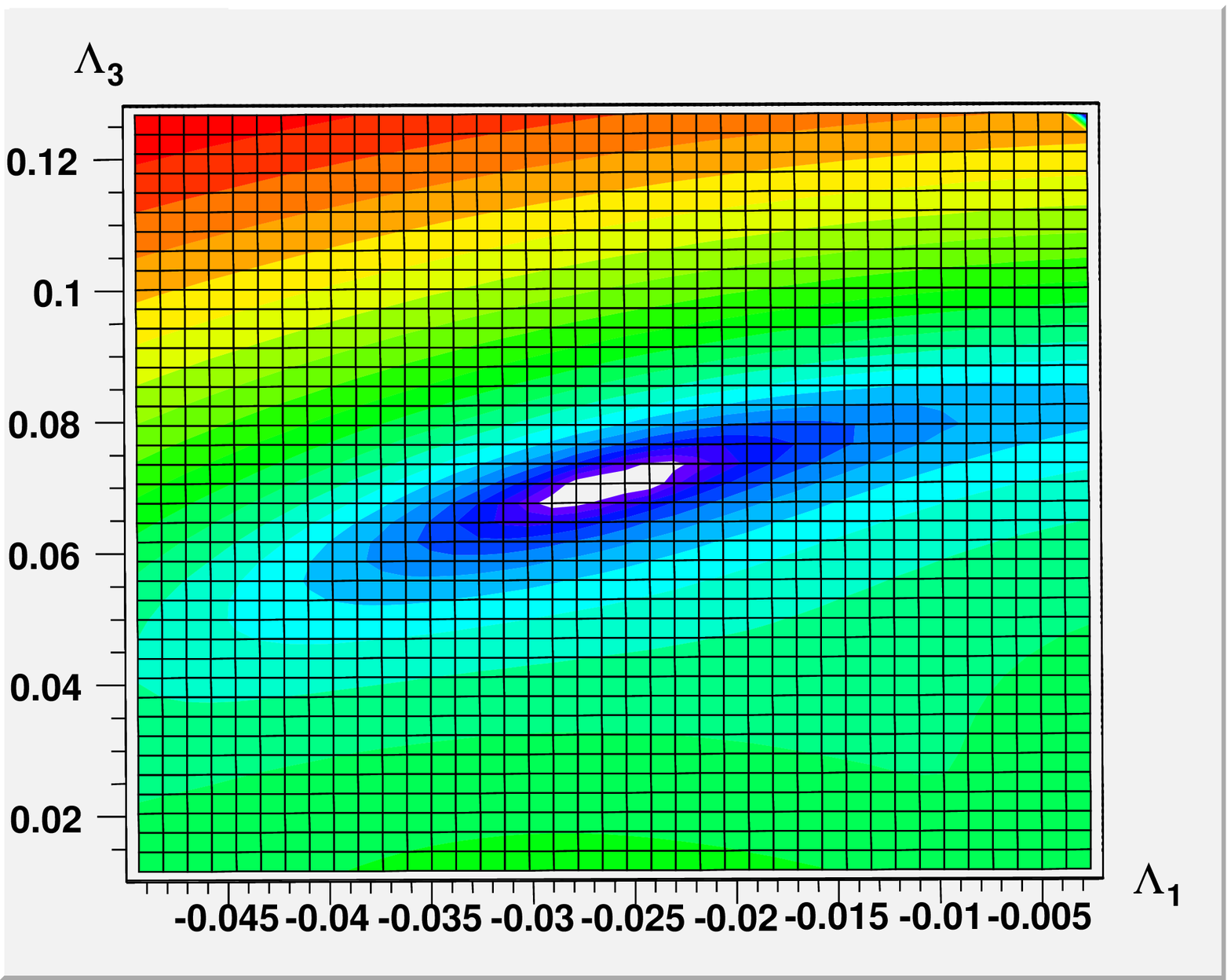,width=0.5\textwidth}}}
\caption{\it Neutrino physics $\chi^2$ as a function of the BRpV parameters 
$\Lambda_1$ and $\Lambda_3$, keeping the rest of the parameters as 
indicated in Table I.
}
\label{L1L3}
\end{figure}

In Fig.~\ref{L1L3} we plot $\chi^2$ as a function of $\Lambda_1$ and 
$\Lambda_3$ with the other parameters as indicated in Table I. The
solar mass squared difference and angle depend strongly on $\Lambda_1$
and $\Lambda_3$ as indicated by the following approximations,
\begin{eqnarray}
\Delta m_{\rm sol}^2 &\approx& A^2\left[
\Lambda_1^2+\frac{\Lambda_3^2}{1+(\epsilon_3/\epsilon_2)^2}\right]^2,
\nonumber\\
\tan^2\theta_{\rm sol} &\approx& 
\frac{\Lambda_1^2}{\Lambda_3^2}\left[1+
\left(\frac{\epsilon_3}{\epsilon_2}\right)^2\right].
\label{approximsol}
\end{eqnarray}
When the $\epsilon$ parameters are kept constant, equal values for the solar
mass are represented by ellipses, while constant values for the solar angle
are represented by straight lines passing through the origen. As with the
previous figure, this geometry can be visualized also in Fig.~\ref{L1L3}.
\subsection{SUSY SU(5) Scenario}
As expressed in Eq.~(\ref{su5numass}) the tree level contribution from BRpV
to the neutrino mass matrix is complemented in our $SU(5)$ supersymmetric
model by a contribution suppressed by one power of the $M_\Sigma$ mass scale.
The dimentionless coefficients $c_i$ are expected to be of order unity, but 
different from each other due to RGE effects. Despite that in Split 
Supersymmetric scenarios the light Higgs cannot contribute at one-loop to 
the neutrino mass matrix, this extra $SU(5)$ term is capable to generate a 
solar mass. 

\begin{table}
\begin{center}
\caption{SU(5) SS and neutrino mass matrix parameters.}
\bigskip
\begin{minipage}[t]{0.8\textwidth}
\begin{tabular}{cccc}
\hline
Parameter & Solution I & Solution II & Units \\
\hline \hline
$\tan\beta$ & 10 & 10 & -  \\
$\mu$ & 450 & 450 & GeV  \\
$M_2$ & 300 & 300 & GeV  \\
$M_1$  & 150 & 150 & GeV  \\
$M_\Sigma$  & $9\times10^{15}$ & $5\times10^{15}$ & GeV  \\
\hline 
$A$ & $-1.7\times10^3$ & $-1.7\times10^3$ & eV/GeV${}^2$ \\
$C$ & $6.7\times10^{-3}$ & $1.2\times10^{-2}$ & eV  \\
\hline \hline \label{tab:SU5abc}
\end{tabular}
\end{minipage}
\end{center}
\end{table}
Keeping the low energy supersymmetric parameters equal to their values in 
the examples shown for Partial Split Supersymmetry in the previous section, 
we look for solutions in the case of SU(5) Split Susy with BRpV. 
In Table III we show two solutions
for two different values of the scale $M_\Sigma$. The resulting neutrino mass
coefficients $A$ and $C$ are also shown in the same Table. The $A$ coefficient
is independent of the mass scale $M_\Sigma$, but $C$ is inversely proportional
to it. 

\begin{table}
\begin{center}
\caption{SU(5) BRpV parameters and neutrino observables.}
\bigskip
\begin{minipage}[t]{0.8\textwidth}
\begin{tabular}{cccc}
\hline
Parameter & Solution I & Solution II & Units \\
\hline \hline
$c_1$ & 0.62 & 0.51 & -  \\
$c_2$ & -0.52 & -1.49 & -  \\
$c_3$ & 0.85 & 1.38 & -  \\
$\lambda_1$  & 0.0008 & 0.0015 & GeV  \\
$\lambda_2$  & -0.0037 & -0.0016 & GeV  \\
$\lambda_3$  & -0.0038 & -0.0011 & GeV  \\
\hline $\Delta m_{atm}^2$ & 2.4$\times10^{-3}$ 
& 2.6$\times10^{-3}$ & eV${}^2$ \\
$\Delta m_{sol}^2$ & 8.2$\times10^{-5}$ & 8.3$\times10^{-5}$ & eV${}^2$ \\
$\tan^2\theta_{atm}$ & 1.02 & 1.00 & - \\
$\tan^2\theta_{sol}$ & 0.45 & 0.50 & - \\
$\tan^2\theta_{13}$ & 0.026 & 0.049 & - \\
$m_{ee}$ & 0.004 & 0.005 & eV \\
\hline \hline \label{tab:SU5sol}
\end{tabular}
\end{minipage}
\end{center}
\end{table}
In Solution I, with a high value for $M_\Sigma=9\times10^{15}$ GeV, the $A$
term in the neutrino mass matrix dominates over the $C$ term, such that the
atmospheric mass comes mainly from $A\lambda_i\lambda_j$, and the smallness 
of the reactor angle is achieved with a small value for $\lambda_1$. The solar
mass is generated with the $Cc_ic_j$ term, with the $c_i$ of order unity.

In Solution II we lower the value for $M_\Sigma=5\times10^{15}$ GeV, reversing
the situation. Now the $Cc_ic_j$ term dominates generating the atmospheric
mass. Since we look for solutions with $c_i$ of order one (we accept 
$0.5<c_i<1.5$), the value of $\tan^2\theta_{13}$ grows to values close to its 
experimental upper bound. In this way, lower values of $M_\Sigma$ are severely
restricted. In this solution the solar mass is generated by the 
$A\lambda_i\lambda_j$ term.

\section{Summary}
\label{conclusions}

We have studied in detail the possibility to describe 
the neutrino masses and mixing angles in the context 
of split supersymmetric scenarios where the sfermions 
and/or Higgses are very heavy. We have considered all 
relevant contributions to the neutrino mass matrix up 
to one-loop level coming from the R-parity violating 
interactions, showing the importance of the Higgs 
one-loop corrections in the case of Partial Split SUSY, 
where only the sfermions are very heavy. We have found new 
contributions in the context of the minimal supersymmetric 
SU(5) which can help us to generate the neutrino masses 
in agreement with the experiments in the SPLIT SUSY scenario.

\begin{acknowledgments}
{\small The work of M.A.D. was partly founded by Conicyt grant 1060629 and
Conicyt and Banco Mundial grant ``Anillo Centro de Estudios Subat\'omicos''.
The work of P.F.P. has been supported by 
{\em Funda\c{c}\~{a}o para a Ci\^{e}ncia e a
Tecnologia} (FCT, Portugal) through the project 
CFTP, POCTI-SFA-2-777 and a fellowship under 
project POCTI/FNU/44409/2002. Also P.F.P. has been 
supported in part by the U.S. Department of Energy
contract No. DE-FG02-08ER41531 and in part by 
the Wisconsin Alumni Research Foundation.
We would like to thank E.~J.~Chun and G.~Senjanovi\'c for a discussion. 
P.~F.~P thanks the Instituto de Fisica Corpuscular (IFIC) in Valencia 
for hospitality. The work by C.~M. was partly supported by Conicyt 
fellowship, and by ALFA-EC in the framework of HELEN program. 
C.~M. would like to thank CERN for their hospitality.}  
\end{acknowledgments}
\appendix
\section{Gauge and Goldstone Boson Loops in Split Susy}
In this appendix we show the properties of the gauge boson one-loop 
contributions to the neutrino mass matrix.
\subsection{Z and Neutral Goldstone Boson Loops}
In $Z$ loops the fermionic sum in Eq.~(\ref{DMgeneral}) is over neutral
fermions $F^0_k$, of which only the neutralinos are relevant. There is
no bosonic sum since only $Z$ contributes. 
\begin{center}
\vspace{-50pt} \hfill \\
\begin{picture}(200,120)(0,23) 
%
%
\ArrowLine(20,50)(80,50)
\Text(50,60)[]{$\nu_j$}
\ArrowArcn(110,50)(30,180,0)
\Text(110,12)[]{$Z$}
\PhotonArc(110,50)(30,180,0){2}{14.5}
\Text(110,95)[]{$\chi^0_k$}
\ArrowLine(140,50)(200,50)
\Text(170,60)[]{$\nu_i$}
\end{picture}
\vspace{30pt} \hfill \\
\end{center}
\vspace{-10pt}
The coupling $G^Z_{ijk}$ is
equal to
\begin{equation}
G^Z_{ijk}=-2(O_{Ljk}^{nnz}O_{Rki}^{nnz}+O_{Rjk}^{nnz}O_{Lki}^{nnz}),
\label{GZnn}
\end{equation}
where the coupling of a $Z$ boson to two neutral fermions 
is \cite{Rosiek:1995kg},
\begin{center}
\vspace{-50pt} \hfill \\
\begin{picture}(110,90)(0,23) 
\Photon(10,25)(50,25){3}{4}
\ArrowLine(50,25)(78,53)
\ArrowLine(78,-3)(50,25)
\Text(10,35)[]{$Z$}
\Text(90,55)[]{$F_i^0$}
\Text(90,-5)[]{$F_j^0$}
\end{picture}
$
=\,i\,\Big[O^{nnz}_{Lij}{{(1-\gamma_5)}\over2}+
O^{nnz}_{Rij}{{(1+\gamma_5)}\over2}\Big]
$
\vspace{30pt} \hfill \\
\end{center}
\vspace{10pt}
with
\begin{equation}
O^{nnz}_{Lij}=-(O^{nnz}_{Rij})^*
\,,\qquad
O^{nnz}_{Rij}=-\frac{g}{2c_W}\left({\cal N}_{i4}^*{\cal N}_{j4}
-{\cal N}_{i3}^*{\cal N}_{j3}-\sum_{a=1}^3
{\cal N}_{ia+4}^*{\cal N}_{ja+4}\right).
\label{OZnn}
\end{equation}
The matrix ${\cal N}$ diagonalizes the $7\times7$ neutrino/neutralino 
mass matrix, giving non-negative eigenvalues. 
Without including the final rotation on the neutrino 
sector, it can be approximated in the following 
way~\cite{Hirsch:2000ef}:
\begin{equation}
{\cal N}\approx\left[\begin{array}{cc}
N    & N\xi^T \\
-\xi & 1
\end{array}\right],
\label{approxN}
\end{equation}
where $N$ diagonalizes the $4\times4$ neutralino mass sub-matrix. The 
parameters $\xi$ are defined by
\begin{eqnarray}
\xi_{i1}&=&\frac{\tilde g'_d \, \mu M_2}{2\det{M_{\chi^0}}}\Lambda_i,
\nonumber\\
\xi_{i2}&=&-\frac{\tilde g_d \, \mu M_1}{2\det{M_{\chi^0}}}\Lambda_i,
\label{xideff}\\
\xi_{i3}&=&\frac{v_u}{4\det{M_{\chi^0}}}\left(
M_1\tilde g_u \tilde g_d + M_2\tilde g'_u\tilde g'_d 
\right)\Lambda_i - \frac{\epsilon_i}{\mu},
\nonumber\\
\xi_{i4}&=&-\frac{v_d}{4\det{M_{\chi^0}}}\left(
M_1\tilde g^2_d + M_2\tilde g'^2_d\right)\Lambda_i.
\nonumber
\end{eqnarray}
For notational brevity we define  the $\xi_i$ parameters as: 
$\lambda_i\xi_1=\xi_{i1}$, $\lambda_i\xi_2=\xi_{i2}$, 
$\lambda_i\xi_3-\epsilon_i/\mu=\xi_{i3}$, and $\lambda_i\xi_4=\xi_{i4}$.
The couplings in Eq.~(\ref{OZnn}) can be approximated with the help of 
Eq.~(\ref{approxN}) to
\begin{equation}
O_{Rik}^{\nu\chi z}\approx \frac{g}{2c_W}\left(
2N_{k4}\xi_{i4}+N_{k1}\xi_{i1}+N_{k2}\xi_{i2}
\right),
\label{appORnnz}
\end{equation}
where $i$ labels the three neutrinos and $k$ labels the four neutralinos.
Considering Eq.~(\ref{xideff}) we conclude,
\begin{equation}
\Delta\Pi_{ij}^{Z}=A^Z\lambda_i\lambda_j,
\end{equation}
with
\begin{equation}
A^Z=-\frac{g^2}{16\pi^2c_W^2}\sum^4_{k=1}\left(
2N_{k4}\xi_4+N_{k1}\xi_1+N_{k2}\xi_2\right)^2 m_{\chi_k^0}
B_0(0;m_{\chi_k^0}^2,m_Z^2).
\end{equation}
This contribution is only a renormalization of the tree level mass 
matrix which it does not break its symmetry, i.e., it does not generate 
mass to all neutrinos. 

There is an extra contribution to $A^Z$ dependent on the gauge parameter
$\xi$. This is canceled by the following loops involving the neutral
Goldstone boson,
\begin{center}
\vspace{-50pt} \hfill \\
\begin{picture}(400,120)(0,23) 
%
%
\ArrowLine(20,50)(80,50)
\Text(50,60)[]{$\nu_j$}
\ArrowArcn(110,50)(30,180,0)
\Text(110,13)[]{$G^0$}
\DashCArc(110,50)(30,180,0){3}
\Text(110,95)[]{$\chi^0_k$}
\ArrowLine(140,50)(200,50)
\Text(170,60)[]{$\nu_i$}
%
%
\Text(235,50)[]{$+$}
\Text(300,60)[]{$\nu_j$}
\ArrowLine(270,50)(330,50)
\Text(360,60)[]{$\nu_i$}
\ArrowLine(330,50)(390,50)
\DashLine(330,50)(330,80){3}
\DashCArc(330,100)(20,180,0){3}
\DashCArc(330,100)(20,360,180){3}
\Text(360,100)[]{$G^0$}
\Text(340,65)[]{$S_k$}
\end{picture}
\vspace{30pt} \hfill \\
\end{center}
\vspace{-10pt}
as demonstrated in Ref.~\cite{Hirsch:2000ef}.
\subsection{W and Charged Goldstone Boson Loops}
In $W$ loops the fermionic sum in Eq.~(\ref{DMgeneral}) is over charged
fermions $F^+_k$, of which only the charginos are relevant. There is
no bosonic sum since only $W$ contributes. 
\begin{center}
\vspace{-50pt} \hfill \\
\begin{picture}(200,120)(0,23) 
%
%
\ArrowLine(20,50)(80,50)
\Text(50,60)[]{$\nu_j$}
\ArrowArcn(110,50)(30,180,0)
\Text(110,11)[]{$W$}
\PhotonArc(110,50)(30,180,0){2}{14.5}
\Text(110,93)[]{$\chi^\pm_k$}
\ArrowLine(140,50)(200,50)
\Text(170,60)[]{$\nu_i$}
\end{picture}
\vspace{30pt} \hfill \\
\end{center}
\vspace{-10pt}
The coupling $G^W_{ijk}$ is
equal to
\begin{equation}
G^W_{ijk}=-4(O_{Ljk}^{ncw}O_{Rik}^{ncw}+O_{Rjk}^{ncw}O_{Lik}^{ncw}),
\label{GWnc}
\end{equation}
where the coupling of a $W$ boson to two fermions is
\begin{center}
\vspace{-50pt} \hfill \\
\begin{picture}(110,90)(0,23) 
\Photon(10,25)(50,25){3}{4}
\ArrowLine(50,25)(78,53)
\ArrowLine(78,-3)(50,25)
\Text(10,35)[]{$W$}
\Text(90,55)[]{$F_i^0$}
\Text(90,-5)[]{$F_j^+$}
\end{picture}
$
=\,i\,\Big[O^{ncw}_{Lij}{{(1-\gamma_5)}\over2}+
O^{ncw}_{Rij}{{(1+\gamma_5)}\over2}\Big]
$
\vspace{30pt} \hfill \\
\end{center}
\vspace{10pt}
with
\begin{eqnarray}
O^{ncw}_{Lij}&=&-g\left({\cal N}_{i2}^*\,{\cal U}_{j1}
+\frac{1}{\sqrt{2}}{\cal N}_{i3}^*\,{\cal U}_{j2}
+\frac{1}{\sqrt{2}}\sum_{a=1}^3{\cal N}_{ia+4}^*\,{\cal U}_{ja+2}
\right),
\nonumber\\
O^{ncw}_{Rij}&=&-g\left({\cal N}_{i2}{\cal V}_{j1}^*
-\frac{1}{\sqrt{2}}{\cal N}_{i4}{\cal V}_{j2}^*\right).
\label{OWnn}
\end{eqnarray}
The ${\cal U}$ and ${\cal V}$ matrices diagonalize the $5\times5$ 
chargino/charged lepton mass matrix, and can be approximated 
to~\cite{Hirsch:2000ef}
\begin{equation}
{\cal U}\approx\left[\begin{array}{cc} U & U\xi_L^T \\ -\xi_L & 1 
\end{array}\right]
\,,\qquad
{\cal V}\approx\left[\begin{array}{cc} V & 0 \\ 0 & 1 \end{array}\right],
\label{appUV}
\end{equation}
where $U$ and $V$ diagonalize the $2\times2$ chargino sub-matrix. The
parameters $\xi_L$ are
\begin{equation}
\xi_L^{i1}=\frac{\tilde g_d}{\sqrt{2}\det{M_{\chi^+}}}\Lambda_i
\,,\qquad
\xi_L^{i2}=-\frac{\tilde g_u\tilde g_d v_u}{2\mu\det{M_{\chi^+}}}
\Lambda_i-\frac{\epsilon_i}{\mu}
\,,
\label{xiLapp}
\end{equation}
with 
\begin{equation}
\det{M_{\chi^+}}=\mu M_2- \frac{1}{2} \tilde g_u \tilde g_d v_u v_d,
\end{equation}
and similarly to what we did in the previous subsection, we define the
parameters $\xi^L_j$, $j=1,2$, with the relations: 
$\xi_L^{i1}=\xi^L_1\lambda_i$
and $\xi_L^{i2}=\xi^L_2\lambda_i-\epsilon_i/\mu$.
The couplings in Eq.~(\ref{OWnn}) can be approximated to
\begin{eqnarray}
O_{Rij}^{\nu\chi w} &\approx& g\left(
V_{j1}^*\xi_{i2}-\frac{1}{\sqrt{2}}V_{j2}^*\xi_{i4}\right),
\nonumber\\
O_{Lij}^{\nu\chi w} &\approx& g\left(
U_{j1}\xi_{i2}-\frac{1}{\sqrt{2}}U_{j2}\left[\xi_L^{i2}-\xi_{i3}\right]-
\frac{1}{\sqrt{2}}U_{j1}\xi_L^{i1}\right),
\label{OROLapp}
\end{eqnarray}
where $i$ labels the three neutrinos and $j$ labels the two charginos.
Similarly to what happened with the $Z$ contributions, the $W$ contribution
depends only on the $\lambda_i$:
\begin{equation}
\Delta\Pi_{ij}^{W}=A^W\lambda_i\lambda_j,
\end{equation}
with
\begin{equation}
A^W=\frac{g^2}{2\pi^2}\sum^2_{k=1}\left[
U_{k1}\xi_2-\frac{U_{k2}}{\sqrt{2}}\left(\xi_2^L-\xi_3\right)
+\frac{U_{k1}}{\sqrt{2}}\xi_1^L
\right]\left(
V_{k1}\xi_2-\frac{V_{k2}}{\sqrt{2}}\xi_4
\right)
m_{\chi_k^+}B_0(0;m_{\chi_k^+}^2,m_W^2).
\end{equation}
Adding to the tree level contribution without changing the symmetry. 
Therefore the $W$ and $Z$ loops do not help us to generate mass to 
all neutrinos.

As for the case of $A^Z$, there is an extra contribution to $A^W$ 
dependent on the gauge parameter $\xi$. This is canceled by loops 
involving the charged Goldstone boson,
\begin{center}
\vspace{-50pt} \hfill \\
\begin{picture}(400,120)(0,23) 
%
%
\ArrowLine(20,50)(80,50)
\Text(50,60)[]{$\nu_j$}
\ArrowArcn(110,50)(30,180,0)
\Text(110,13)[]{$G^\pm$}
\DashCArc(110,50)(30,180,0){3}
\Text(110,95)[]{$\chi^\pm_k$}
\ArrowLine(140,50)(200,50)
\Text(170,60)[]{$\nu_i$}
%
%
\Text(235,50)[]{$+$}
\Text(300,60)[]{$\nu_j$}
\ArrowLine(270,50)(330,50)
\Text(360,60)[]{$\nu_i$}
\ArrowLine(330,50)(390,50)
\DashLine(330,50)(330,80){3}
\DashCArc(330,100)(20,180,0){3}
\DashCArc(330,100)(20,360,180){3}
\Text(360,100)[]{$G^\pm$}
\Text(340,65)[]{$S_k$}
\end{picture}
\vspace{30pt} \hfill \\
\end{center}
\vspace{-10pt}
The rest of the tadpoles form a gauge invariant set, and renormalize
the vacuum expectation values~\cite{Hirsch:2000ef}.
\subsection{Charged Higgs Boson Loops}
The last loops we consider are the ones which include a charged 
scalar and a charged fermion. The loop is represented by the 
following graph,
\begin{center}
\vspace{-50pt} \hfill \\
\begin{picture}(200,120)(0,23) 
%
%
\ArrowLine(20,50)(80,50)
\Text(50,60)[]{$\nu_j$}
\ArrowArcn(110,50)(30,180,0)
\Text(110,13)[]{$H^+$}
\DashCArc(110,50)(30,180,0){3}
\Text(112,94)[]{$\chi^+_k$}
\ArrowLine(140,50)(200,50)
\Text(170,60)[]{$\nu_i$}
\end{picture}
\vspace{30pt} \hfill \\
\end{center}
\vspace{-10pt}
where the $G$ factor in Eq.~(\ref{DMgeneral}) is,
\begin{equation}
G^{s+}_{ijkr}=(O_{Ljkr}^{ncs}O_{Lkir}^{cns}+O_{Rjkr}^{ncs}O_{Rkir}^{cns}).
\label{Gspnn}
\end{equation}
The relevant coupling above the scale $\tilde m$ is the charged 
scalar couplings to a charged and a neutral fermion. It is given 
by,
\begin{center}
\vspace{-50pt} \hfill \\
\begin{picture}(110,90)(0,23) 
\DashLine(10,25)(50,25){3}
\ArrowLine(50,25)(78,53)
\ArrowLine(78,-3)(50,25)
\Text(10,35)[]{$S^+_k$}
\Text(90,55)[]{$F_i^+$}
\Text(90,-5)[]{$F_j^0$}
\end{picture}
$
=\,\Big[O^{cns}_{Lijk}\frac{(1-\gamma_5)}{2}+
O^{cns}_{Rijk}\frac{(1+\gamma_5)}{2}\Big]
$
\vspace{30pt} \hfill \\
\end{center}
\vspace{10pt}
where the $O^{cns}_L$ and $O^{cns}_R$ couplings are,
\begin{eqnarray}
O^{cns}_{Lijk}&=&
h_\tau R^{S^+}_{k1}{\cal N}^*_{j7}{\cal V}^*_{i5}-
R^{S^+}_{k2}\bigg(\frac{g}{\sqrt{2}}{\cal N}^*_{j2}{\cal V}^*_{i2}+
\frac{g'}{\sqrt{2}}{\cal N}^*_{j1}{\cal V}^*_{i2}+
g{\cal N}^*_{j4}{\cal V}^*_{i1}\bigg)
\nonumber\\ && \quad -
h_\tau R^{S^+}_{k5}{\cal N}^*_{j3}{\cal V}^*_{i5}-
\sqrt{2} g' R^{S^+}_{k\,\ell+5}{\cal N}^*_{j1}{\cal V}^*_{i\,\ell+2},
\label{chHcoupl}\\
O^{cns}_{Rijk}&=&
R^{S^+}_{k1}\bigg(\frac{g}{\sqrt{2}}{\cal N}_{j2}{\cal U}_{i2}+
\frac{g'}{\sqrt{2}}{\cal N}_{j1}{\cal U}_{i2}-
g{\cal N}_{j3}{\cal U}_{i1}\bigg)+
h_\tau R^{S^+}_{k8}\Big({\cal N}_{j7}{\cal U}_{i2}-
{\cal N}_{j3}{\cal U}_{i5}\Big)
\nonumber\\ && \quad +
R^{S^+}_{k\,\ell+2}
\bigg(\frac{g}{\sqrt{2}}{\cal N}_{j2}{\cal U}_{i\,\ell+2}+
\frac{g'}{\sqrt{2}}{\cal N}_{j1}{\cal U}_{i\,\ell+2}-
g{\cal N}_{j\,\ell+4}{\cal U}_{i1}\bigg),
\nonumber
\end{eqnarray}
with $O^{cns}_{Lijk}=O^{ncs\,*}_{Rjik}$ and $O^{cns}_{Rijk}=O^{ncs\,*}_{Ljik}$.
The fields $S^+_k$ are eight linear combinations of charged Higgs bosons 
and charged sleptons, whose mass matrix in the 
$(H_d^+,H_u^+,\tilde\ell^+_{Lj},\tilde\ell^+_{Rj})$ basis is in Appendix B.
This mass matrix is diagonalized in PSSusy by the rotation,
\begin{equation}
\left(\begin{matrix}
G^+ \cr H^+ \cr \tilde l^+_{Li} \cr \tilde l^+_{Ri}
\end{matrix}\right)=
\left(\begin{matrix}
 c_\beta & s_\beta & -s_L^j & 0 \cr
-s_\beta & c_\beta & -t_L^j & 0 \cr
c_\beta s_L^i-s_\beta t_L^i & s_\beta s_L^i+c_\beta t_L^i & \delta_{ij} & 0 \cr
0 & 0 & 0 & \delta_{ij} \cr
\end{matrix}\right)
\left(\begin{matrix}
H^+_d \cr H^+_u \cr \tilde\ell^+_{Lj} \cr \tilde\ell^+_{Rj}
\end{matrix}\right).
\end{equation}
An expression for the mixing angles $s_L^i$ and $t_L^i$ above the
scale $\tilde m$ can be found in the Appendix B. Comparing the 
supersymmetric lagrangian above he scale $\tilde m$ in Eq.~(\ref{susyRpV})
with the terms of the PSSusy lagrangian in Eq.~(\ref{LSS2HDMRpV}) we
find the following matching conditions,
\begin{equation}
s_L^i(\tilde m)=b_i(\tilde m) s_\beta
\,;\qquad
t_L^i(\tilde m)=b_i(\tilde m) c_\beta,
\label{sL_p}
\end{equation}
where $s_L^i(\tilde m)$ represents the amount of slepton $\tilde L_i$
present in the charged Goldstone boson $G^+$, and analogously with 
$t_L^i(\tilde m)$ for the low energy charged Higgs $H^+$. In the limit 
where the sleptonic fields have a very large mass,
\begin{equation}
s_L^i\rightarrow s_\beta\frac{v_i}{v_u}\,,\qquad
t_L^i\rightarrow c_\beta\frac{v_i}{v_u},
\end{equation}
indicating that $b_i = v_i/v_u$ in agreement with the CP-even and CP-odd
cases. Now we make an expansion of the couplings in Eq.~(\ref{chHcoupl})
and we find 
\begin{eqnarray}
O^{\nu\chi h+}_{Lik}&=& c_\beta
\bigg(\frac{g}{\sqrt{2}}\xi_{i2}V^*_{k2}+
\frac{g'}{\sqrt{2}}\xi_{i1}V^*_{k2}+
g\xi_{i4}V^*_{k1}\bigg),
\nonumber\\
O^{\nu\chi h+}_{Rik}&=&s_\beta\bigg(
\frac{g}{\sqrt{2}}\xi_{i2}U_{k2}+
\frac{g'}{\sqrt{2}}\xi_{i1}U_{k2}+
g\xi_{i3}U_{k1}\bigg)+gt_L^iU_{k1},
\end{eqnarray}
and isolating the terms proportional to $\epsilon_i$, using 
Eq.~(\ref{sL_p}), we write,
\begin{equation}
O_{Lik}^{\nu\chi h+}=\widetilde O_{Lk}^{\nu\chi h+}\Lambda_i
\,,\qquad
O_{Rik}^{\nu\chi h+}=\widetilde O_{Rk}^{\nu\chi h+}\Lambda_i-
\frac{1}{\mu s_\beta} g U_{k1} \epsilon_i,
\label{OnuXh3}
\end{equation}
where we have defined,
\begin{eqnarray}
\widetilde O^{\nu\chi h+}_{Lk} &=& c_\beta
\bigg(\frac{g}{\sqrt{2}}\xi_2V^*_{k2}+
\frac{g'}{\sqrt{2}}\xi_1V^*_{k2}+
g\xi_4V^*_{k1}\bigg),
\nonumber\\
\widetilde O^{\nu\chi h+}_{Rk} &=& s_\beta\bigg(
\frac{g}{\sqrt{2}}\xi_2U_{k2}+
\frac{g'}{\sqrt{2}}\xi_1U_{k2}+
g\xi_3U_{k1}\bigg)+gU_{k1}\frac{c_\beta}{\mu v_u}.
\end{eqnarray}
Finally, the charged Higgs contribution to the neutrino mass matrix is,
\begin{eqnarray}
\Delta\Pi_{ij}^{h+}&=&-\frac{1}{16\pi^2}\sum_{k=1}^2
\widetilde O^{\nu\chi h+}_{Lk}\Big[
2\widetilde O^{\nu\chi h+}_{Rk}\Lambda_i\Lambda_j-
\frac{g U_{k1}}{\mu s_\beta}\big(
\Lambda_i\epsilon_j+\Lambda_j\epsilon_i \big)\Big]
m_{\chi_k^+}B_0(0;m_{\chi_k^+}^2,m_{H^+}^2).
\label{4loops}
\end{eqnarray}
Note that there is no $\epsilon_i\epsilon_j$ term.
\section{Higgs Slepton Sector}
Here we give details on the Higgs Slepton mass matrices and approximations
in the case when the slepton masses are much heavier that the Higgs masses.
\subsection{CP-even Higgs Sneutrino Mixing}
The CP-even Higgs and sneutrino fields mix to form a set of five neutral 
mass eigenstates $S^0_i$. We organize the mass terms in the lagrangian
in the following way,
\begin{equation}
{\cal L} \owns -\frac{1}{2}\Big[\phi^0_d,\phi^0_u,\tilde\ell^0_{si}
\Big] \, {\bf M}_{S^0}^2 \left[
\begin{matrix}\phi^0_d \cr \phi^0_u \cr \tilde\ell^0_{sj} \end{matrix}
\right].
\end{equation}
The mass matrix is divided into blocks~\cite{Hirsch:2000ef},
\begin{equation}
{\bf M}_{S^0}^2 = \left[
\begin{matrix}
{\bf M}_{S^0hh}^2 &  {\bf M}_{S^0h\widetilde{\nu}}^2 \cr
{\bf M}_{S^0h\widetilde{\nu}}^{2T}&
{\bf M}_{S^0\widetilde{\nu} \widetilde{\nu}}^2  
\end{matrix}
\right].\label{S0}
\end{equation}
The Higgs $2\times2$ sub-matrix is equal to,
\begin{equation}
{\bf M}_{S^0hh}^2= \left[
\begin{matrix}
B_0\mu{{v_u}\over{v_d}}+{\textstyle{\frac{1}{4}}} g_Z^2v_d^2+
\mu \vec\epsilon\cdot \frac{\vec v}{v_d}+{{T_d}\over{v_d}}
& -B_0\mu-{\textstyle{\frac{1}{4}}} g^2_Zv_dv_u  
\cr 
-B_0\mu-{\textstyle{\frac{1}{4}}} g^2_Zv_dv_u  
&
B_0\mu{{v_d}\over{v_u}}+{\textstyle{\frac{1}{4}}} g^2_Zv_u^2-
\vec B_{\epsilon}\cdot {{\vec v}\over{v_u}}+{{T_u}\over{v_u}}
\end{matrix}\right], 
\label{MS0hh}
\end{equation}
where we call $g_Z^2=g^2+g'^2$, and in supergravity models we have
$B_\epsilon^i=B_i\epsilon_i$. In this
matrix we have eliminated the Higgs soft masses using the minimization
conditions of the scalar potential (or tadpole equations) 
\cite{Hirsch:2000ef}. These Higgs tadpole equations at tree level are,
\begin{eqnarray}
T_d&=&\Big(m_{H_d}^2+\mu^2\Big)v_d+v_dD-
\mu\Big(B_0v_u+\vec v\cdot\vec\epsilon\Big),
\nonumber\\
T_u&=&-B_0\mu v_d+\Big(m_{H_u}^2+\mu^2\Big)v_u-v_uD+\vec v\cdot\vec B_\epsilon
+v_u\vec\epsilon^{\,2},
\label{tadpolesUD}
\end{eqnarray}
with $D={\textstyle{\frac{1}{8}}}(g^2+g'^2)(\vec v^{\,2}+v_d^2-v_u^2)$.
At tree level, it is safe to set $T_u=T_d=0$, and if 
we take the R-Parity conserving limit $\epsilon_i,v_i\rightarrow0$, we can 
recognize the CP-even Higgs mass matrix of the MSSM. The $2\times3$ mixing 
sub-matrix is given by,
\begin{equation}
{\bf M}_{S^0h\widetilde\nu}^2= \left[
\begin{matrix}
M_{S^0h_d\tilde\nu_i}^2
\cr
M_{S^0h_u\tilde\nu_i}^2
\end{matrix}\right]
= \left[
\begin{matrix}
-\mu\epsilon_i+{\textstyle{\frac{1}{4}}} g^2_Zv_dv_i  
\cr
B_\epsilon^i-{\textstyle{\frac{1}{4}}} g^2_Zv_uv_i  
\end{matrix}\right],
\label{HsnuMix}
\end{equation}
which vanishes in the R-Parity conserving limit. Finally, the sneutrino 
sub-matrix is given by,
\begin{equation}
\left({\bf M}_{S^0\widetilde\nu \widetilde\nu}^2\right)_{ij}
=\left(M^2_{Li}+D\right)\delta_{ij} 
+{\textstyle{\frac{1}{4}}} g^2_Zv_iv_j + \epsilon_i \epsilon_j,
\end{equation}
where we have not yet used the corresponding tadpole equations, and we have
assumed that the sneutrino soft mass matrix is diagonal. The sneutrino
tadpole equations are given by, 
\begin{equation}
T_i=v_iD+\epsilon_i\left(-\mu v_d+\vec v\cdot\vec\epsilon\right)+
v_uB_\epsilon^i+v_iM^2_{Li}.
\label{eq:tadpoles_i}
\end{equation}
It is clear from this equation that if the sneutrino vev's are zero,
$\mu\epsilon_i=B_\epsilon^i v_u/v_d$, and therefore, the mixing between
the up and down Higgs fields with the sneutrino fields are related by 
$M_{S^0h_d\tilde\nu}^2=-\tan\beta M_{S^0h_u\tilde\nu}^2$. Of course, this last
relation is not valid if the sneutrino vev's are not zero.

In the case of large slepton masses, the mass matrix in Eq.~(\ref{S0})
is diagonalized in two steps by the rotation matrix,
\begin{equation}
R_{S^0}=
\left(
\begin{matrix}
  1   &   0  &    -s_s^j   \cr
  0   &   1  &    -t_s^j   \cr
s_s^i & t_s^i & \delta_{ij}
\end{matrix}\right)
\left(
\begin{matrix}
-s_\alpha & c_\alpha &     0      \cr
 c_\alpha & s_\alpha &     0      \cr
    0    &     0   & \delta_{ij}
\end{matrix}\right),
\end{equation}
with the mixing angles at the scale $\widetilde m$ satisfying,
\begin{equation}
s_s^i=\frac{-s_\alpha M_{S^0h_d\tilde\nu_i}^2+c_\alpha M_{S^0h_u\tilde\nu_i}^2}
{M^2_{L_i}-m_h^2}
\,,\qquad
t_s^i=\frac{c_\alpha M_{S^0h_d\tilde\nu_i}^2+s_\alpha M_{S^0h_u\tilde\nu_i}^2}
{M^2_{L_i}-m_H^2},
\label{si}
\end{equation}
where the Higgs masses can be neglected in front of the slepton masses
in this approximation. From Eq.~(\ref{HsnuMix}) we find the following 
limits for large slepton masses,
\begin{equation}
s_s^i\longrightarrow-c_\alpha\frac{v_i}{v_u}\,,\qquad
t_s^i\longrightarrow-s_\alpha\frac{v_i}{v_u},
\end{equation}
which links the smallness of the Higgs-sneutrino mixing needed for
neutrino physics, with the smallness of the sneutrino vevs.
\subsection{CP-odd Higgs-Sneutrino Mixing}
The CP-odd Higgs bosons and sneutrinos mix to form a set of five CP-odd
scalars, whose mass terms in the Lagrangian are,
\begin{equation}
{\cal L} \owns -\frac{1}{2}\Big[\varphi^0_d,\varphi^0_u,\tilde\ell^0_{pi}
\Big] \, {\bf M}_{P^0}^2 \left[
\begin{matrix}\varphi^0_d \cr \varphi^0_u \cr \tilde\ell^0_{pj}
\end{matrix}\right],
\end{equation}
where the $5\times 5$ mass matrix we decompose in the following blocks,
\begin{equation}
{\bf M}_{P^0}^2 = \left[
\begin{matrix}
{\bf M}_{Phh}^2 &  {\bf M}_{Ph\widetilde{\nu}}^2 \cr
{\bf M}_{Ph\widetilde{\nu}}^{2T}&
{\bf M}_{P\widetilde{\nu} \widetilde{\nu}}^2  
\end{matrix}
\right].
\end{equation}
The Higgs sector is given by the $2\times 2$ mass sub-matrix,
\begin{equation}
{\bf M}_{Phh}^2= \left[
\begin{matrix}
B_0\mu{{v_u}\over{v_d}}+
\mu \vec\epsilon\cdot \frac{\vec v}{v_d}+{{T_d}\over{v_d}}
& B_0\mu
\cr 
B_0\mu 
&
B_0\mu{{v_d}\over{v_u}}-
\vec B_{\epsilon}\cdot {{\vec v}\over{v_u}}+{{T_u}\over{v_u}}
\end{matrix}\right], 
\end{equation}
where the tadpoles $T_u$ and $T_d$ are defined in Eq.~(\ref{tadpolesUD}). In
the R-Parity conserving limit we reproduce the CP-odd mass matrix in the MSSM.
The higgs-sneutrino mixing is given by the $2\times 3$ matrix,
\begin{equation}
{\bf M}_{Ph\widetilde\nu}^2= \left[
\begin{matrix}
M_{Ph_d\tilde\nu_i}^2
\cr
M_{Ph_u\tilde\nu_i}^2
\end{matrix}\right]= \left[
\begin{matrix}
-\mu\epsilon_i
\cr
-B_\epsilon^i
\end{matrix}\right],
\label{HsnuMix2}
\end{equation}
which vanishes in the R-Parity conserving limit. Finally, the sneutrino
$3\times 3$ mass matrix is,
\begin{equation}
\left({\bf M}_{P\widetilde\nu \widetilde\nu}^2\right)_{ij}
=\left(M^2_{Li}+D\right)\delta_{ij} + \epsilon_i \epsilon_j,
\end{equation}
where we have assumed diagonal soft slepton mass parameters.

If slepton masses are very large, the $5\times 5$ mass matrix can be
diagonalized with the following rotations,
\begin{equation}
R_{P^0}=
\left(
\begin{matrix}
  1   &   0  &    -s_p^j   \cr
  0   &   1  &    -t_p^j   \cr
s_p^i & t_p^i & \delta_{ij}
\end{matrix}\right)
\left(
\begin{matrix}
-c_\beta & s_\beta &     0      \cr
 s_\beta & c_\beta &     0      \cr
    0    &     0   & \delta_{ij}
\end{matrix}\right),
\end{equation}
with the mixing angles $s_p^i$ and $t_p^i$ satisfying at the scale 
$\widetilde m$, 
\begin{equation}
s_p^i=\frac{-c_\beta M_{Ph_d\tilde\nu_i}^2+s_\beta M_{Ph_u\tilde\nu_i}^2}
{M^2_{L_i}-m_G^2}
\,,\qquad
t_p^i=\frac{s_\beta M_{Ph_d\tilde\nu_i}^2+c_\beta M_{Ph_u\tilde\nu_i}^2}
{M^2_{L_i}-m_A^2},
\label{si2}
\end{equation}
and the Higgs masses $m_G^2$ and $m_A^2$ negligible in front of the 
slepton masses. Using Eqs.~(\ref{HsnuMix2}) and (\ref{si2}) we find
the following mixing angles in the limit of large slepton masses,
\begin{equation}
s_p^i\longrightarrow s_\beta\frac{v_i}{v_u}\,,\qquad
t_p^i\longrightarrow c_\beta\frac{v_i}{v_u},
\end{equation}
also proportional to the sneutrino vacuum expectation values.
\subsection{Charged Higgs Slepton Mixing}
The charged Higgs boson and slepton fields mix to form a set of eight 
charged eigenstates $S^+_i$, whose mass terms in the lagrangian are 
organized according to,
\begin{equation}
{\cal L} \owns -\Big[H^-_d,H^-_u,\tilde\ell_{Li}^-,\tilde\ell_{Ri}^-
\Big] \, {\bf M}_{S^+}^2 \left[
\begin{matrix} H^+_d \cr H^+_u \cr \tilde\ell_{Li}^+ \cr \tilde\ell_{Ri}^+ 
\end{matrix}
\right].
\end{equation}
The $8\times 8$ mass matrix is written as,
\begin{equation}
{\bf M}_{S^+}^2 = \left[
\begin{matrix}
{\bf M}_{S^+hh}^2 &  {\bf M}_{S^+h\widetilde{\ell}}^2 \cr
{\bf M}_{S^+h\widetilde{\ell}}^{2T}&
{\bf M}_{S^+\widetilde{\ell} \widetilde{\ell}}^2  
\end{matrix}
\right],
\end{equation}
with the following charged Higgs boson $2\times 2$ block,
\begin{equation}
{\bf M}_{S^+hh}^2= \left[
\begin{matrix}
B_0\mu{{v_u}\over{v_d}}+\mu \vec\epsilon\cdot \frac{\vec v}{v_d}
+\frac{1}{4}g^2(v_u^2-\vec v^2)+\frac{1}{2}h_{\ell k}^2v_k^2
+{{T_d}\over{v_d}}
& B_0\mu+\frac{1}{4}g^2v_dv_u
\cr 
B_0\mu+\frac{1}{4}g^2v_dv_u
&
B_0\mu{{v_d}\over{v_u}}-
\vec B_{\epsilon}\cdot {{\vec v}\over{v_u}}
+\frac{1}{4}g^2(v_d^2+\vec v^2)+{{T_u}\over{v_u}}
\end{matrix}\right]. 
\end{equation}
This mass matrix reduces to the charged Higgs mass matrix of the MSSM when
the BRpV parameters are taken equal to zero. Mixing between charged Higgs
bosons and left and right charged sleptons appear through terms in the 
following $2\times6$ block,
\begin{equation}
{\bf M}_{S^+h\widetilde\ell}^2= \left[
\begin{matrix}
M_{S^+h_d\tilde\ell_{Li}}^2 & M_{S^+h_d\tilde\ell_{Ri}}^2 
\cr
M_{S^+h_u\tilde\ell_{Li}}^2 & M_{S^+h_u\tilde\ell_{Ri}}^2
\end{matrix}\right]= \left[
\begin{matrix}
-\mu\epsilon_i+\frac{1}{4}g^2v_dv_i-\frac{1}{2}h_{\ell i}^2v_dv_i
&
-\frac{1}{\sqrt{2}}h_{\ell i}^2v_u\epsilon_i
-\frac{1}{\sqrt{2}}A_{\ell i}v_i
\cr
-B_\epsilon^i+\frac{1}{4}g^2v_uv_i
&
-\frac{1}{\sqrt{2}}h_{\ell i}(\mu v_i+\epsilon_i v_d)
\end{matrix}\right],
\label{HsnuMix3}
\end{equation}
which as expected vanishes in the R-Parity conserving limit. The charged 
slepton sub-matrix is further divided into left and right slepton sectors,
\begin{equation}
{\bf M}_{S^+\widetilde{\ell} \widetilde{\ell}}^2 = \left[
\begin{matrix}
{\bf M}_{LL}^2 &  {\bf M}_{LR}^2 \cr
{\bf M}_{LR}^{2T}&
{\bf M}_{RR}^2  
\end{matrix}
\right],
\end{equation}
which are given by the following expressions,
\begin{eqnarray}
M^2_{LL}&=&\left[M_{L_i}^2+\frac{1}{8}(g^2-g'^2)(v_u^2-v_d^2-\vec v^2)
+\frac{1}{2}h_{\ell i}^2\,v_d^2 \right]\delta_{ij}
+\frac{1}{4}g^2v_iv_j+\epsilon_i\epsilon_j,
\nonumber\\
M^2_{LR}&=&\frac{1}{\sqrt{2}}\,(v_dA_{\ell i}-\mu v_uh_{\ell i})\,\delta_{ij},
\\
M^2_{RR}&=&\left[M_{R_i}^2+\frac{1}{4}g'^2(v_u^2-v_d^2-\vec v^2)
+\frac{1}{2}h_{\ell i}^2\,(v_d^2+\vec v^2) \right]\delta_{ij}.
\nonumber
\end{eqnarray}
Slepton soft mass parameters are taken diagonal, and the MSSM expressions 
are recovered when we make $\epsilon_i=v_i=0$. As before, if slepton soft
masses are large, a diagonalization can be accomplished by the rotations,
\begin{equation}
R_{S^+}=
\left(
\begin{matrix}
  1   &   0  &    -s_{L}^j  & -s_{R}^j   \cr
  0   &   1  &    -t_{L}^j  & -t_{R}^j   \cr
s_{L}^i & t_{L}^i & \delta_{ij} & 0       \cr
s_{R}^i & t_{R}^i &   0     & \delta_{ij}
\end{matrix}\right)
\left(
\begin{matrix}
 c_\beta & s_\beta &     0   & 0   \cr
-s_\beta & c_\beta &     0   & 0   \cr
    0    &     0   & \delta_{ij} & 0 \cr
    0    &     0   & 0      & \delta_{ij}
\end{matrix}\right),
\end{equation}
with the following mixing angles at the scale $\widetilde m$,
\begin{eqnarray}
s_L^i=\frac{c_\beta M_{S^+h_d\tilde\ell_{Li}}^2+s_\beta M_{S^+h_u\tilde\ell_{Li}}^2}
{M^2_{L_i}-m_{H^+}^2}
\,,&\qquad&
t_L^i=\frac{-s_\beta M_{S^+h_d\tilde\ell_{Li}}^2+c_\beta M_{S^+h_u\tilde\ell_{Li}}^2}
{M^2_{L_i}-m_{G^+}^2},
\nonumber\\
s_R^i=\frac{c_\beta M_{S^+h_d\tilde\ell_{Ri}}^2+s_\beta M_{S^+h_u\tilde\ell_{Ri}}^2}
{M^2_{R_i}-m_{H^+}^2}
\,,&\qquad&
t_R^i=\frac{-s_\beta M_{S^+h_d\tilde\ell_{Ri}}^2+c_\beta M_{S^+h_u\tilde\ell_{Ri}}^2}
{M^2_{R_i}-m_{G^+}^2}.
\label{si3}
\end{eqnarray}
When slepton masses are very large, the right mixing angles vanish while the
left mixing angles are proportional to the slepton vevs,
\begin{equation}
s_L^i\longrightarrow s_\beta\frac{v_i}{v_u}\,,\qquad
t_L^i\longrightarrow c_\beta\frac{v_i}{v_u}\,,\qquad
s_R^i\longrightarrow 0\,,\qquad
t_R^i\longrightarrow 0,
\end{equation}
in a similar way as the previous two cases.

\end{document}